\documentstyle[12pt]{article}
\newcommand{\Hg}[1]{{$^{19#1}$Hg}}
\newcommand{\Pb}[1]{{$^{19#1}$Pb}}
\newcommand{\Tl}[1]{{$^{19#1}$Tl}}
\newcommand{\Au}[1]{{$^{19#1}$Au}}

\newcommand{\be}{\begin{equation}}
\newcommand{\ee}{\end{equation}}
\newcommand{\ba}{\begin{array}}
\newcommand{\ea}{\end{array}}
\newcommand{\bc}{\begin{center}}
\newcommand{\ec}{\end{center}}

\newcommand{\q}{\quad}
\newcommand{\usd}{{\frac{1}{2}}}
\newcommand{\usq}{{\frac{1}{4}}}

\newcommand{\wphig}{{\langle\Psi\vert}}
\newcommand{\wphid}{{\vert\Psi\rangle}}
\newcommand{\wph}[1]{{\phi_{#1}({\mbox{\boldmath$r$}})}}
\newcommand{\wpr}[1]{{{\varphi}_{#1}({\mbox{\boldmath$r$}})}}
\newcommand{\wprp}[1]{{{\varphi}_{#1}({\mbox{\boldmath$r'$}})}}

\newcommand{\HH}{{\hat H}}
\newcommand{\JJ}{{\hat J}_x}
\newcommand{\NN}{{\hat N}}
\newcommand{\OO}{{\hat O}}
\newcommand{\TT}{{\hat T}}
\newcommand{\VV}{{\hat V}}

\newcommand{\HDN}{{\Delta\NN}}

\newcommand{\mn}[1]{{\langle\HDN^{#1}\rangle}}
\newcommand{\mnu}{{\mn{}}}
\newcommand{\mnd}{{\mn{2}}}
\newcommand{\mnt}{{\mn{3}}}
\newcommand{\mnq}{{\mn{4}}}
\newcommand{\mhun}{{\langle\HH\HDN\rangle}}
\newcommand{\mhde}{{\langle\HH(\HDN^2-\mnd)\rangle}}

\newcommand{\tr}[1]{{\hbox{tr}\left[#1\right]}}

\newcommand{\CE}{{\cal E}}

\newcommand{\CW}{{\cal W}}

\newcommand{\CHT}{{\hat{\cal{T}}}}

\newcommand{\HJ}{{J}}
\newcommand{\HN}{{\delta}}
\newcommand{\HT}{{T}}
\newcommand{\HV}{{V}}

\newcommand{\UZ}{{U}}
\newcommand{\VZ}{{V}}
\newcommand{\EZ}{{E}}

\newcommand{\dd}{{\Delta}}
\newcommand{\epp}{e^\omega}
\newcommand{\eqp}{E^\omega}
\newcommand{\ggg}{{\gamma}}
\newcommand{\hh}{{h}}
\newcommand{\kk}{{\kappa}}
\newcommand{\lll}{{\lambda}}

\newcommand{\rr}{{\rho}}
\newcommand{\oo}{{\omega}}
\newcommand{\vv}{{n}}
\newcommand{\xx}{{\chi}}

\newcommand{\vmf}{{v_{HF}\{\xx\}}}
\newcommand{\vpa}{{v_P\{\ggg\kk\}}}

\newcommand{\nun}{{n_1}}
\newcommand{\nde}{{n_2}}

\newcommand{\rrr}{{\rho({\mbox{\boldmath$r$}},{\mbox{\boldmath$r$}}\,')}}

\newcommand{\hi}{{\hat{\imath}}}
\newcommand{\hj}{{\hat{\jmath}}}

\newcommand{\hl}{{\hat{l}}}

\newcommand{\bfdummy}{}
\newcommand{\fpu}{{\bfdummy u}}
\newcommand{\ftu}{{\tilde{\bfdummy u}}}
\newcommand{\fpv}{{\bfdummy v}}
\newcommand{\ftv}{{\tilde{\bfdummy v}}}
\newcommand{\fpr}{{\bfdummy r}}
\newcommand{\ftr}{{\tilde{\bfdummy r}}}
\newcommand{\fph}{{\bfdummy h}}
\newcommand{\fth}{{\tilde{\bfdummy h}}}
\newcommand{\fpk}{{\bfdummy k}}
\newcommand{\fpd}{{\bfdummy d}}

\newcommand{\ca}[1]{{ a_{#1}^+ }}
\newcommand{\da}[1]{{ a_{#1} }}
\newcommand{\cb}[1]{{ b_{#1}^+ }}
\newcommand{\db}[1]{{ b_{#1} }}

\newcommand{\jde}{{{\cal J}^{(2)}}}
\newcommand{\CEP}[1]{{\CE^P_{#1}}}
\newcommand{\skm}{{SkM$^*$}}
\newcommand{\umi}{{$\hbar^2$MeV$^{-1}$}}
\newcommand{\ho}{{$\hbar\oo$}}
\newcommand{\qp}{{Q_c}}

\newcommand{\nsu}   [1]{$\nu[770]^{#1}_{1/2}$}
\newcommand{\nsd}   [1]{$\nu[761]^{#1}_{3/2}$}
\newcommand{\nczco} [1]{$\nu[505]^{#1}_{11/2}$}
\newcommand{\nsqzu} [1]{$\nu[640]^{#1}_{1/2}$}
\newcommand{\nsqdt} [1]{$\nu[642]^{#1}_{3/2}$}
\newcommand{\ncudc} [1]{$\nu[512]^{#1}_{5/2}$}
\newcommand{\nsdqn} [1]{$\nu[624]^{#1}_{9/2}$}
\newcommand{\nst}   [1]{$\nu[752]^{#1}_{5/2}$}
\newcommand{\psd}   [1]{$\pi[651]^{#1}_{3/2}$}
\newcommand{\pctzu} [1]{$\pi[530]^{#1}_{1/2}$}
\newcommand{\pctdt} [1]{$\pi[532]^{#1}_{3/2}$}
\newcommand{\pst}   [1]{$\pi[642]^{#1}_{5/2}$}

\newcommand{\pquuu} [1]{$\pi[411]^{#1}_{1/2}$}

\newcommand{\eppmax}{\epp_{\mbox{\scriptsize{max}}}}
\newcommand{\llld}{{\lambda_2}}
\newcommand{\xgt}{g}
\newcommand{\xGt}{G}

\title{
Superdeformed rotational bands in the\\
Mercury region; A Cranked
\\Skyrme-Hartree-Fock-Bogoliubov study
}
\author{
B. Gall\\
Centre de Spectrom\'etrie Nucl\'eaire et de Spectrom\'etrie de Masse \\
Bt. 104, F-91405 Orsay Campus, France
\and
P. Bonche\\
Service de Physique Th\'eorique, DSM, CE Saclay\\
F-91191 Gif-sur-Yvette Cedex, France
\and
J. Dobaczewski\\
Institute of Theoretical Physics, Warsaw University, Ho\.za 69\\
PL-00681, Warsaw, Poland
\and
H. Flocard\\
Division de Physique Th\'eorique\thanks{
Unit\'e de recherche des Universit\'es Paris VI et Paris XI,
associ\'ee au Centre National de la Recherche Scientifique
}, Institut de Physique Nucl\'eaire\\
F-91406 Orsay Cedex, France
\and
P.-H. Heenen\thanks{
Directeur de Recherches FNRS
}\\
Physique Nucl\'eaire Th\'eorique, Universit\'e Libre de Bruxelles,\\
CP229, B-1050, Bruxelles, Belgium
}
\begin{document}

\maketitle

\centerline{Preprint IPN-TH 93-66}

\newpage

\begin{abstract}
A  study of rotational properties of the
ground superdeformed bands in \Hg{0}, \Hg{2}, \Hg{4}, and \Pb{4}
is presented.
We use the cranked Hartree-Fock-Bogoliubov method with
the {\skm} parametrization
of the
Skyrme force in the particle-hole channel and a seniority interaction
in the pairing channel.
An approximate particle number projection is performed by means of
the Lipkin-Nogami prescription.
We analyze the proton and neutron
quasiparticle routhians in connection with the
present information on about
thirty presently observed superdeformed bands in nuclei
close neighbours of \Hg{2}.

\end{abstract}

\section{Introduction}

With the recent evolution in $\gamma$-ray multidetectors, the
experimental knowledge of superdeformed (SD)
bands in nuclei of the mercury region
is rapidly growing. Since the initial discovery
of a SD rotational
band in \Hg{1} \cite{MOO89}, more than forty bands have been observed
for  isotopes of
Au, Hg, Tl and Pb (see
refs.~\cite{LIA92,LAU92,VOK93,FAL93a,JOY93a,%
HAN93,GAL93a,JOY93c,PIL93,DUP93b,CED93}
and the earlier papers cited in the compilation of experimental
data \cite{HAN92}).
The anchor point of the region appears
to be the nucleus \Hg{2} which corresponds to the SD magic numbers:
$Z$=80 and $N$=112.

Although  superdeformation has been
predicted by many models (for a review see ref.~\cite{ABE90a}),
the spectrum of approaches used up to now
to study high spin properties of the SD states is still
rather limited. The theoretical analyses have been mostly
done in terms of the
cranked Nilsson-Strutinsky \cite{TSA70} method, either with
the modified harmonic oscillator (MHONS),
e.g. ref.~\cite{ABE90b}, or with
the Wood-Saxon (WSNS), e.g.
ref.~\cite{SAT91,CHA90,SAT93}, potentials.
The methods relying on a mean field derived from a
microscopic nucleon-nucleon force have been applied only in two
cases. For \Hg{2}, \Hg{4} and \Pb{4} the cranked Hartree-Fock
equations have been solved \cite{CHE92} and
in \Hg{2} the Bohr hamiltonian approach
with the collective potential and mass parameters calculated at
zero spin has been used \cite{GIR92}.

In view of their general good
quality at describing zero-spin properties,
the latter methods should be more extensively tested at
high spins.
Such methods are free from geometric
assumptions concerning a detailed parametrization
of the nuclear surface, and therefore they
seem to be more apt at describing polarization
phenomena that may appear
in the dynamical behaviour of a rotating nucleus.

As compared to the other major region of superdeformation
(around $A$=150) the high spin properties of nuclei
in the \Hg{2} region seem
to be different because of the
presence of large pairing correlations.
In particular, the generally observed increase of the $\jde$ moment of inertia
is interpreted \cite{RIL90,DRI91} as resulting both from
the alignment of intruder orbitals and from a gradual disappearance
of pairing correlations.
The relative importance of these two effects
requires further microscopic investigations.
There exist also uncertainties concerning the predictions of
crossing frequencies, the determination of
pairing strengths and the
evaluation of the relative importance of
dynamical
pairing correlations  compared to
those described by
the static pairing formalism which is
usually sufficient to account for zero spin properties.

The present work is devoted to a presentation of
the cranked Hartree-Fock-Bogoliubov (HFB) method and to a discussion
of its ability to address such issues.
The next section contains a short description of our theoretical
approach. In particular, we present a method of solution of the
cranked HFB equations with the Skyrme interaction in the coordinate
representation.
Pairing correlations are described by a seniority interaction and we
briefly describe how we implement an approximate variation
after projection by means of the Lipkin-Nogami
approach \cite{LIP60,NOG64,PRA73} (HFBLN).

As a first application we use the cranked HFB method to study the
SD bands in nuclei around \Hg{2}.
Section 3 contains a presentation of our results.
First we investigate properties of the ground SD
bands in \Hg{0}, \Hg{2}, \Hg{4} and \Pb{4}.
Then, we show that the properties of SD
bands observed in other odd and even nuclei in this region are
compatible with our spectra of quasiparticle
routhians. The last section presents our conclusions.

\section{The HFBLN method}
\subsection{THE TWO-BASIS METHOD}

First, we review the cranked HFB method {\cite{RIN80}}
as much as required to introduce our notation.
Given the hamiltonian $\HH$
which in a single particle basis $\{\ca{i},\,\da{i}, (i=1,\ldots,N)\}$
has the form
\be\ba{rcl}\label{Meq10}
\HH&=&\TT+\VV \q , \\
\TT&=&\sum_{ij}\,\HT_{ij}\ca{i}\da{j} \q, \\
\VV&=&\usq\sum_{ijkl}\,\HV_{ijkl}\ca{i}\ca{j}\da{l}\da{k} \q, \\
\ea\ee
this method determines the quasiparticle
vacuum state $\wphid$
which minimizes
the energy
\be\label{Meq11}
\CE = \wphig\HH\wphid
\ee
subject to the constraints
$\wphig\NN\wphid=N_0$
and
$\wphig\JJ\wphid=J_0$.
The average value of the particle-number operator $\NN$ is thus
constrained to the desired number of nucleons $N_0$
(separately for neutrons and protons)
and that of
the cranking operator $\JJ$ to the desired value of angular
momentum\footnote{
In general, $J_0$ is related to the
angular momentum $J$ and its projection $K$ on the symmetry axis
by $J_0=\sqrt{J(J+1)-K^2}$.
Since we are mostly interested in high spin properties and $K$=0 bands
we take $J_0\approx J$.
}
$J_0$.
The collective-cranking operator $\JJ$
is equal to the projection
of the angular-momentum vector $\hat{\mbox{\boldmath$J$~}}$ on the axis
perpendicular to the largest-elongation axis.

The determination of $\wphid$ amounts to
finding the unitary matrix
$\CW$ which relates the particle creation and annihilation
operators $\{\ca{i},\,\da{i}\}$
to the quasiparticle operators $\{\cb{i},\,\db{i}\}$
such that the $\db{}$'s are destruction operators
associated with the quasiparticle vacuum
$\wphid$ ($\,\db{i}\wphid=0,\,i=1,\ldots,N$):
\be\label{Meq20}
\CW=\left(
\ba{cc}
\UZ&\VZ^*\\
\VZ&\UZ^*
\ea
\right)\q;\q
\left(\ba{c}
\cb{i}\\ \db{i}
\ea\right)
=\CW^+
\left(\ba{c}
\ca{i}\\ \da{i}
\ea\right)\q.
\ee
The minimization of the total routhian $\CE^\oo$,
\be\label{Meq21}
\CE^\oo =  \CE-\lll\wphig\NN\wphid-\oo\wphig\JJ\wphid,
\ee
leads to the HFB eigenvalue problem
for the $2N\times2N$ matrix:
\be\label{Meq40}
\left(\ba{cc}
\hh&\dd\\-\dd^*&-\hh^*
\ea\right)
\left(\ba{c}\UZ \\ \VZ \ea\right)=
\left(\ba{c}\UZ \\ \VZ \ea\right)\EZ\q,
\ee
which gives two $N\times N$ matrices $\UZ$ and $\VZ$ as eigenvectors
and the diagonal matrix $\EZ$ of $N$ eigenvalues which are the
quasiparticle routhians (qpr), $\EZ_{ij}$=$\delta_{ij}\eqp_i$.
The complete HFB spectrum is composed of pairs of
opposite eigenvalues \cite{RIN80}
and in Eq.~(\ref{Meq40}),
only one eigenvalue from each pair should be  selected.
The Hartree-Fock hamiltonian $\hh$
and the pairing potential $\dd$
are constructed from the density matrix $\rr$ and the pairing
tensor $\kk$:
\be\ba{rcl}\label{Meq50}
\rr&=&\VZ^*\VZ^T\q, \\
\kk&=&\VZ^*\UZ^T\q, \\
\ea\ee
according to
\be\ba{rcl}\label{Meq60}
\hh_{ij}&=&\HT_{ij}+\sum_{kl}\,\HV_{ikjl}\,\rr_{lk}
                          -\lll\HN_{ij}\,-\oo\HJ_{ij}\q, \\
\dd_{ij}&=&\usd\,\sum_{kl}\,\HV_{ijkl}\,\kk_{kl}\q. \\
\ea\ee

In order to find a self-consistent solution of the HFB equation
(\ref{Meq40}) we have adopted a method which relies
on the
simultaneous use of two different bases
of single-particle states. The first one
is the basis of functions $\wph{i}$ which converge to the
eigenfunctions of the Hartree-Fock hamiltonian
$\hh$ and the second one is the canonical basis, i.e., the
basis of eigenstates of the density matrix $\rr$. The iteration procedure
is performed in the following way.

Suppose we start at some point
with a set of orthogonal functions $\wph{i}$ and with the hamiltonian
$\hh$ in the coordinate representation. First, we perform
one imaginary-time-step evolution of the set $\wph{i}$ using
$\hh$ as an evolution operator. The imaginary-time evolution
\cite{DAV80}
has already become a standard technique in the HF coordinate-space
calculations. It ensures that as iterations proceed, the set
$\wph{i}$ converges to the eigenfunctions of $\hh$
with the single-particle routhians $\epp_i$
as eigenvalues,
$h_{ij}$=$\delta_{ij}(\epp_i$$-$$\lll)$.

Second, we
calculate the matrix elements  $\dd_{ij}$ of
the  pairing potential
and the diagonal matrix elements $\hh_{ii}$
of the Hartree-Fock hamiltonian
in the
basis of the new (imaginary-time evolved) wave functions
$\wph{i}$. As discussed above the matrix $\hh_{ij}$ converges
to the diagonal matrix

and therefore we can safely disregard its off-diagonal
matrix elements during the iteration.
On the other hand, for \ho$\neq$0 all the matrix
elements $\dd_{ij}$ are
in general nonzero
both during the iteration and {\em after} the convergence is obtained.

Third, we solve Eq.~(\ref{Meq40}) to obtain the matrices
$\UZ$ and $\VZ$, the quasiparticle
routhians $\eqp_i$
as well as the matrices $\rr$ and $\kk$ by using
Eq.~(\ref{Meq50}).
A diagonalization of the density matrix,
\be\label{Meq51}
\rr_{kj} = \sum_i \vv_i W_{ki} W^+_{ij} \q ,
\ee
provides us with the occupation
coefficients $\vv_i$ and with the unitary transformation $W$
which relates the basis $\wph{i}$ to
the canonical-basis wave functions
$\wpr{i}$=$\sum_j W_{ji}\wph{j}$.

{}Finally, we construct the HFB density matrix in the
coordinate space
\be\label{Meq70}
\rrr=\sum_i\,\vv_i\,\wpr{i}\wprp{i}^*\q,
\ee
which is given as a single sum of products of the wave functions
$\wpr{i}$. This fact is crucial for a numerically effective
calculation of local
Skyrme densities on the spatial mesh and then of various
terms in the Hartree-Fock hamiltonian.
In fact, at this point the density matrix has a form identical
to that used in the HF+BCS method although it corresponds to an
unrestricted HFB variation.
At the expense of handling two sets of single-particle wave
functions we are thus able to solve the HFB equations by using
techniques which have already been well developed for
the HF+BCS method.

{}For our purpose, the two-basis method has
several advantages.
{}First of all it allows
us to use the existing techniques to obtain solutions of
the cranked HF equations
associated with a Skyrme interaction
directly on a three dimensional mesh.
These have been presented in
refs.~\cite{BON87a,BON91a} and
have been applied to study
the high spin properties of $^{24}$Mg and $^{80}$Sr.
Because they do not rely on an expansion on a basis, they can
describe with the {\em same} good accuracy the very different shapes
that a rotating nucleus can take (some examples are shown in Fig.~7
of \cite{BON91a}).

Moreover, a separate treatment of the particle-hole
and particle-particle channels becomes possible.
Although there exist effective
forces which have been shown to describe reasonably well
both channels of interaction simultaneously,\footnote{
For instance the SkP Skyrme interaction \cite{DOB84} or
the earlier and mathematically sounder global
effective interaction proposed by Gogny
\cite{GOG73,GOG75,DEC75,DEC80,BER84}}
we believe that the present
knowledge of the pairing effective interactions is not as
sophisticated as that concerning the mean-field.
This is especially true
as far as its properties relevant to
situations of time-reversal symmetry breaking such as encountered
in high spins physics
are concerned. Treating in two
steps the particle and the quasiparticle spaces, permits
us to evaluate the gap matrix in the particle basis $\wph{i}$
with a different interaction than the one used in the cranked
HF equation.

Our two-basis method provides also
a more natural scheme for limiting the phase space than
the truncation generated by a fixed basis.
Because we are solving the HFB problem (\ref{Meq40})
directly in the cranked HF basis,
its dimension $N$
can be chosen much smaller than that needed for
the accurate description of the HF orbitals $\wph{i}$.
Similarly as in the phenomenological approaches
\cite{CWI80} and based on our
former experience, we limit it to orbitals
for which the particle routhians
$\epp_i$ do not exceed a given value $\eppmax$.
This value should be much larger than $\lll$+$\dd$,
where $\dd$ is the physical pairing gap.
The results presented below have been obtained with
$\eppmax$=$\lll$+5~MeV.
This leads to a dimension of the HFB matrix of the order
of hundred, compared to typical dimensions of the cranked HF equations
of the order of several thousand.

By applying the two-basis method we have circumvented the
difficulties associated with the fact that
the HFB hamiltonian has a spectrum unbounded
from below in the coordinate
representation. Therefore, a direct application of any variational
method (e.g., of the imaginary-time method) is impossible.
When spherical symmetry is enforced, the HFB problem becomes
one-dimensional and its solutions
can be obtained \cite{DOB84} by explicitly solving a differential
Schr\"odinger-like equation.
However this route seems to be difficult
to take for deformed systems which
require a three-dimensional treatment.
In practice, we use the double representation
inherent to the two-basis method, to
filter  those  states
from the unbounded HFB spectrum
which are the most physically important.

\subsection{THE SENIORITY PAIRING INTERACTION IN THE CRANKED BASIS}

In the present study, we use the simplest possible pairing force:
the seniority interaction. This force being compatible with
the signature symmetry, we can separate the phase
space in two subspaces of dimensions $\nun$ and $\nde$
($\nun+\nde=N$) associated with opposite signatures. In the following,
we denote respectively with latin  (i.e. $i$)
and  hat-topped latin indices (i.e. $\hi$)
states in the positive and negative signature subspaces.
Then the non-zero matrix elements of the seniority forces
are
\be\label{Seq1}
\HV_{i\hj k\hl}=-\xGt\langle\hj\vert\CHT\vert i\rangle
\langle\hl\vert\CHT\vert k\rangle\q,
\ee
plus all those obtained from the antisymmetry conditions.
In (\ref{Seq1}), $\CHT$ stands for the time-reversal operator.
This expression corresponds to the lowest-order
truncation of the pairing matrix elements of a short range
force. The pairing strength $G$ is
parametrized for neutrons (protons) as
$\xGt_N=\xgt_N/(11+N)$ ($\xGt_Z=\xgt_Z/(11+Z)$),
as was done in our preceding studies.
We discuss later
how to select
the values of the two parameters $\xgt_N$ and $\xgt_Z$.

According to a well established procedure
\cite{GOO74}, we consider the
matrices $\UZ$ and $\VZ$ in the following block form
\be\label{Seq2}
\UZ=\left(\ba{cc}
\fpu&0 \\ 0&\ftu^*
\ea\right)
\q;\q
\VZ=\left(\ba{cc}
0&\ftv^* \\ -\fpv &0
\ea\right)\q,
\ee
in which the first block has dimension $\nun$ and the second
dimension $\nde$.
This leads to simplifications for $\rr$ and $\kk$:
\be\label{Seq3}
\rr=\left(\ba{cc}
\fpr&0 \\ 0&\ftr^*
\ea\right)
\q;\q
\kk=\left(\ba{cc}
0&\fpk \\ -\fpk^T &0
\ea\right)\q;\q
\left\{\ba{l}
\fpr=\ftv\ftv^+\\
\ftr=\fpv\fpv^+\\
\fpk=\ftv\ftu
\ea\right.\q,
\ee
and for $\hh$ and $\dd$:
\be\label{Seq4}
\hh=\left(\ba{cc}
\fph&0 \\ 0&\fth^*
\ea\right)
\q;\q
\dd=\left(\ba{cc}
0&\fpd \\ -\fpd^T &0
\ea\right)
\q,
\ee
where the expressions for $\fph$, $\fth$ and $\fpd$ can be obtained
by inserting Eqs.~(\ref{Seq3}) in Eqs.(\ref{Meq60}).
Then the HFB equations (\ref{Meq40}) reduce to the
following $N$-dimensional
problem:
\be\label{Seq5}
\left(\ba{cc}
\fph & \fpd \\
\fpd^+ & -\fth
\ea\right)
\left(\ba{cc}
\fpu & \ftv \\
-\fpv & \ftu
\ea\right)
=
\left(\ba{cc}
\fpu & \ftv \\
-\fpv & \ftu
\ea\right)
\left(\ba{cc}
\eqp_i\HN_{ij} & 0 \\
 0 & -\eqp_{\hi}\HN_{\hi\hj}
\ea\right)\q.
\ee

This last equation specifies our definition of the
quasiparticle routhians (qpr)
assuming that the eigenvectors of the diagonalization
problem are ordered according to {\em decreasing} eigenvalues
($\nun$ first, $\nde$ last). The first $\nun$ eigenvectors
are associated with the positive signature.
The last $\nde$
eigenvectors are associated with the negative signature
and the corresponding eigenvalues
 are opposite
to the negative-signature qpr's.
In particular, following the discussion
in ref.~\cite{CHU75} concerning the
parity of the number of particles
in the BCS state, we do not consider the signs of the eigenvalues
to decide which qpr's are of a given signature.

In a similar manner, to study
one-quasiparticle or two-quasiparticle states, we
exchange the $\UZ$ and $\VZ$
components of eigenvectors according to the following procedure.
{}First, we choose a given configuration by selecting
the label $i$ of the wave function $\wph{i}$ upon which the quasiparticle
excitation is to be constructed.
Then we scan the $i$th line in the
matrix
$\left(\ba{cc}
\fpu & \ftv \\
-\fpv & \ftu
\ea\right)
$
looking for the largest component
to determine the column (i.e. the qpr) for which the
exchange of the $\UZ$ and $\VZ$ components must be done.
We can also determine whether this particular
quasiparticle excitation is predominantly of the particle
or of the hole type; in the former (latter) case the largest component
is located in the $\fpu$ or $\ftu$ ($\fpv$ or $\ftv$) blocks.
We have found this procedure more stable than that using the
values of qpr's to select quasiparticle excitations.
Indeed, during the course
of the HFB iterations, the displacement of the Fermi level
necessarily associated with the creation of one or two
quasiparticles
makes it difficult to  recognize a
given configuration solely on the basis of the
value of its qpr.

\subsection{THE LIPKIN-NOGAMI PRESCRIPTION}

Although the HFB equations contain most of the
ingredients essential for the understanding
of the physics of fast rotating nuclei, they present a
significant weakness inherent to their mean-field
nature: for specific values of the angular velocity,
they predict phase transitions
which, in principle, should not occur in a finite system
such as the nucleus.
As we will show, such transitions generate
characteristic behaviours in the evolution
of the moment of inertia versus angular velocity
which are incompatible with present data.
They correspond to a sudden breakdown of neutron or
proton pairing
correlations and are associated with a poor description
of particle-number symmetry breaking in situations
of small particle-number fluctuations.
To cure this deficiency while retaining the
quality and the simplicity of a mean-field description,
the best approach would be a variation after projection (VAP)
on correct particle number \cite{RIN80}.
Because it appeared too difficult
to implement it in this first study, we  have adopted
the approximate projection method proposed by Lipkin and
Nogami \cite{LIP60,NOG64,PRA73}.
The associated equations which we recall below
have been tested to provide a good numerical approximation
of the VAP in situations where both the HFB
and the BCS equations
predict a collapse of the pairing correlations.
(See the recent studies of model systems without \cite{ZHE92,DOB93}
and with rotation \cite{MAG93}, and the references cited therein.)

The prescription of Lipkin-Nogami amounts to modify the
energy $\CE$ (or the total routhian $\CE^\oo$ in case of
cranked HFBLN method) by adding the
second-order Kamlah \cite{KAM68}
correction:
\be\label{Leq1}
\CE\rightarrow \CE-\llld\mnd\q,
\ee
where $\langle\OO\rangle$ denotes the expectation
value $\wphig\OO\wphid$.
The coefficient $\llld$ is given by:
\be\label{Leq2}
\llld=\frac{\displaystyle
\mhde-\mhun\mnt/\mnd
}{\displaystyle
\mnq-\mnd^2-\mnt^2/\mnd}\q.
\ee
{}For the HFB state, the moments
of the operator $\HDN=\NN-\mnu$ which appear in the definition
of $\llld$ are given by \cite{FLO93}:
\be\label{Leq3}\left\{\ba{l}
\displaystyle \mnd=2\,\tr{\xx}\\
\displaystyle \mnt=4\,\tr{\ggg\xx}\\
\displaystyle \mnq=3\mnd^2+8\,\tr{\xx(1-6\xx)}
\ea\right.\q;\q
\left\{\ba{l}
\xx=\rr(1-\rr)\\
\ggg=1-2\rr
\ea\right.\q,
\ee
and
\be\label{Leq4}\left\{\ba{rl}
\mhun&=\displaystyle 2\,\tr{\hh\xx}-\Re\,\tr{\dd\kk^*\ggg}\\
\mhde&=\displaystyle 4\,\tr{(\hh\ggg+\vmf)\xx}
-\Re\,\tr{\dd\kk^*(1-8\xx)}\\
&\q\q\q
-\tr{\vpa(\ggg\kk)^*}
\ea\right.\q,
\ee
where the symbol $\tr{a}$ stands
for the trace of the matrix $a$
and $\Re$ denotes the real part.
The mean fields $\vmf$ and $\vpa$ are the HF potential and
the pairing potential
calculated with the densities $\xx$ and $\ggg\kk$, respectively:
\be\label{Leq5}
\vmf_{ij}=\sum_{kl}\,\HV_{ikjl}\xx_{lk}\q;\q
\vpa_{ij}=\usd\sum_{kl}\,\HV_{ijkl}(\ggg\kk)_{kl}\q.
\ee

A consistent application of the Lipkin-Nogami prescription
requires in principle
using the full effective interaction in Eqs.~(\ref{Leq5}).
In the present study we calculate $\llld$ by using only
the seniority pairing interaction.
However, in (\ref{Leq4}) the effects of the Skyrme interaction
are still taken into account by the contributions containing
the HF hamiltonian $\hh$.
This is consistent with all the other applications of the
Lipkin-Nogami prescription performed up to date.
The fact that particle-number
mixing is directly related to the existence of an
interaction in the particle-particle
channel
may be considered as a justification for such a procedure.
Admittedly, further studies of this question are needed.

Assuming that the quasiparticle vacuum $\wphid$ is a
self-consistent solution, one may derive another formula
for $\llld$ which gives the same value as (\ref{Leq2})
 at the end
of the iteration procedure. In fact, this
approach is usually taken in applications,
because it gives
 $\llld$
in terms of the two-body interaction only.
For the case of the seniority pairing interaction such explicit
expressions are given in refs.~\cite{PRA73} and \cite{MAG93}
without and with time-reversal symmetry breaking,
respectively.

The modification of the HFB equations associated with the
Lipkin-Nogami prescription is obtained by a restricted variation of
$\llld\mnd$, namely,
$\llld$ is not varied although its value
is calculated self-consistently using (\ref{Leq2}).
Based on Eq.~(\ref{Leq3}), this leads \cite{PRA73}
to the modification
of the HF hamiltonian, $\hh\rightarrow\hh-2\llld\ggg$,
while the pairing potential $\dd$ is unchanged.
In the present study
we have used this method to study \Hg{2}, \Hg{4} and \Pb{4}.
However, as noted
in ref.~\cite{SAT93}, one may use an equivalent form
of $\mnd$=$2\tr{\kk^+\kk}$, which leads to
$\dd\rightarrow\dd-4\llld\kk$ and leaves $\hh$ unchanged.
In this way, a better numerical stability of the HFBLN equations
can be obtained \cite{SAT93}.
We have also investigated an intermediate
prescription:
\be\label{Leq6}
\hh\rightarrow\hh-\llld\ggg\q;\q\dd\rightarrow\dd-2\llld\kk\q,
\ee
which results from inserting the explicit form of the
particle-number operator ($N_{ij}$=$\delta_{ij}$) only after
the variations over $\rr$ and $\kk$ are performed.
Such a prescription also turns out to be numerically stable
and was used to study \Hg{0}.

\section{Results}

The calculations reported  below have been performed
using the {\skm} pa\-ra\-met\-ri\-za\-tion \cite{BAR82} of the Skyrme
force and the seniority pairing interaction as discussed in
sec.~2.2.
The method of solution
of the cranked HF equations, which requires
breaking of the time-reversal symmetry, has been discussed
for the Skyrme interaction in refs.~\cite{BON87a,BON91a}.
 The single-particle
wave functions are discretized on a spatial rectangular
mesh of points spaced by $\Delta x$=$\Delta y$=$\Delta z$=1fm.
The parity and signature  symmetries are assumed to be
conserved, which allows us to solve the equations by using only
points in space which have positive values of all three coordinates.

Although the code does not enforce it, the
self-consistent solutions that we have obtained
 deviate
 little from axial symmetry: in
all calculations the triaxiality angle $\gamma$ never
exceeds $\pm$0.2$^\circ$.
The average value of the projection of
angular momentum on the axis perpendicular to the symmetry axis $J_0$
is constrained while that of the quadrupole moment
$Q$ is unconstrained. In this way,
the self-consistent quadrupole moment of the SD minimum
is found for each value of the angular momentum
and the evolution of $Q$ is obtained
as function of $J_0$.

We use a
quadratic constraint on the angular momentum and therefore
 we are able to select
values of $J_0$
along the rotational band
 equal to even integers.
The  Lagrange multiplier $\oo$ required to obtain
a given value
of $J_0$ is interpreted as the
angular velocity.
Due to self-consistency the
second moment of inertia $\jde$,  defined as
\be\label{Req1}
\jde=\frac{\partial J_0}{\partial\oo},
\ee
can be expressed
in terms of the first derivative of the energy (\ref{Meq11})
with respect to $\oo$,
\be\label{Req12}
\jde=\frac{1}{\oo}\frac{\partial \CE}{\partial\oo} .
\ee

Neglecting a weak dependence
on $\oo$ of the Lagrange multipliers $\lll$
associated with proton and neutron numbers,
one can also express the angular
momentum as the first derivative of the total routhian
$\CE^\oo$, Eq.~(\ref{Meq21}),
\be\label{Req10}
J_0=-\frac{\partial\CE^\oo}{\partial\oo}.
\ee

Within the same approximation one therefore obtains
an alternate
formula for the second moment of inertia:
\be\label{Req11}
\jde=-\frac{\partial^2\CE^\oo}{\partial\oo^2} .
\ee
Because  formulae
(\ref{Req1}) and (\ref{Req12}) involve a first order
derivative, they are
numerically more stable than (\ref{Req11}).
Moreover,
formula (\ref{Req1}) which uses the
average value of a one-body operator
is easier to handle than
(\ref{Req12}) which depends on the two-body
interaction.
We have found that an accurate determination of $\jde$
requires a high level of convergence.
In order to make connection with the experimental method
for determining the second moment of inertia we have also
calculated $\jde$ according to:
\be\label{Req13}
\jde=\frac{4\hbar^2}{E_{I+2}-2E_I+E_{I-2}} ,
\ee
where $E_I$ is the total energy $\CE$ obtained by constraining
the average value of the cranking angular momentum to even
integers, i.e.,  $J_0$=$I$.

In practice
we have to perform about
200 iterations (or 300 in the crossing regions)
with an imaginary
time-step $\delta t$=0.015$\hbar$/MeV.
Only then do we obtain the second moments of inertia
calculated from (\ref{Req1}) and (\ref{Req13}) equal up to
1\%.

At this point one should stress an
important difference between the interpretation of quasiparticle
routhians (qpr's) in the HF and the Nilsson-Strutinsky (NS) approaches.
In both cases, the value of the angular momentum $J_0$
is the sum of contributions from individual quasiparticle states,
$J_0$=$\sum_i j_i$, where $j_i$ are the individual alignments,
i.e., the average values of the cranking angular momentum $\JJ$
in the quasiparticle states.
Therefore, the second moment of inertia (\ref{Req1}) is also
a sum of contributions from individual quasiparticle states,
equal to the first derivatives of alignments with respect to $\oo$.
On the other hand, the NS individual alignments are equal to the
opposite of the first
derivatives of the NS qpr's, and hence
their contributions to $\jde$ are given by the second derivatives
of qpr's.
This is no longer true for the HFB qpr's because
the HF mean field {\em depends self-consistently} on the angular velocity
(cf. discussion in ref.~\cite{CHE92}). Only the {\em total} values of the
angular momentum $J_0$ and of the second moment of inertia $\jde$
are equal to the opposite of the first and the second derivatives
of the {\em total} routhian,
Eqs.~(\ref{Req10}) and (\ref{Req11}), respectively.
They have, however, no direct relations
with the sums of the first and the second derivatives of qpr's
with respect to $\oo$. This fact should be kept in mind
when analyzing the diagrams of the HFB or HFBLN qpr's. In
particular,
the changes in $\jde$ caused by a quasiparticle excitation
cannot not be simply inferred from these diagrams.

\subsection{EFFECT OF DIFFERENT TREATMENTS OF PAIRING CORRELATIONS}

{}Figure \ref{FIG1} presents a comparison of the second moment
of inertia $\jde$
calculated within the cranked HF, HFB and HFBLN approximations
for the nucleus \Hg{2}.

The HF moment of inertia\footnote{
There is a 2{\umi} reduction of the value of $\jde$ as compared
with the calculations of ref.~\cite{CHE92} which have been
performed with $\Delta x$=0.8fm.
It measures the absolute
inaccuracy of the present calculation (performed with
$\Delta x$=1fm) in replacing the
differential operators of the Skyrme functional
and of the angular momentum by a finite-difference
expression on the spatial mesh.
}
keeps an  almost constant
value ($\approx 115$\umi) versus angular velocity.
A similar result obtained within the WSNS method is reported in
ref.~\cite{RIL90}.
This constancy is in contradiction with the data which
show a steady increase of
$\jde$ from 90 to 135{\umi} over the range
100keV$\le$\ho$\le$450keV. One is therefore led to
conclude that in the mercury region
pairing correlations  play an important role
in the superdeformed high-spin states.
The increase of $\jde$ would then be
at least partially attributed to a
reduction of the superfluidity as the magnitude of the
time-reversal-breaking cranking field grows.

The HFB results obtained with a seniority force $\xgt_N=\xgt_Z$=15.5MeV
do indeed display an increase of $\jde$ (Fig.~\ref{FIG1})
which is however too rapid at low rotation frequencies.
Despite a value
of $\jde$(0)$\equiv$$\jde$(\ho=0) which is
lower than indicated by experiment, the
moment of inertia becomes larger than
the experimental values in less than ten
units of angular momentum. In addition, the HFB curve presents
two sudden drops at \ho=250keV and 350keV.
These two accidents in the $\jde$
curve are due to the collapses of
the proton and neutron pairing correlations.
This is shown by Fig.~\ref{FIG2}
which displays the evolution versus {\ho} of
the proton and neutron pairing contributions to the total energy
defined by:
\be\label{Req2}
\CEP{\tau}(\hbar\oo)=-\tr{\dd_\tau\kk_\tau^*}\q,
\ee
where the index $\tau$ refers to neutrons ($N$) or protons ($Z$).
The fact that the proton pairing disappears first and
the neutron pairing later
is related to the magnitude of the pairing correlations
at \ho=0 which are weaker for protons than for neutrons.
This in turn reflect the corresponding sizes of the SD shell
effects.

{}Fig.~\ref{FIG3}
gives the evolution of the neutron and proton
single-particle energies versus quadrupole deformation in the
neighbourhood of the SD well.
One sees that for the {\skm} interaction,
the proton magic gap at
$Z$=80 is large and dominates the neutron shell effect at
$N$=112 which occurs at a slightly larger deformation.
WSNS calculations also find that this proton
gap in the single-particle spectrum is
larger than the neutron gap.
As a consequence, at \ho=0 the proton pairing
correlations are weaker
than neutron ones and tend therefore to
disappear first.
Once the pairing correlations have vanished, the HFB
moment of inertia
$\jde$ becomes identical to that of the HF method.

The introduction of
an approximate treatment of particle number projection
by the Lipkin-Nogami method modifies
the behaviour of $\jde$ and $\CEP{\tau}$
as function of \ho.
Because the projection always leads to an increase
of pairing correlations (cf. ref~\cite{MOL92}), we had
to lower the pairing strengths to
$\xgt_N=\xgt_Z=14$MeV so as to obtain
a moment of inertia close to that found with the HFB method
at \ho=0.
{}Fig.~\ref{FIG2} shows that despite this reduction,
the proton pairing energy $\CEP{Z}$(0) is significantly
increased over the HFB value while $\CEP{N}$(0) is
much less affected.
This reflects another characteristics of the
particle-number projection
which induces larger relative modifications whenever
the pairing correlations
are weak, i.e., when the particle shell effects are large.

Within the HFBLN method
the pairing energies decrease smoothly
with the angular velocity (Fig.~\ref{FIG2}).
The spurious sudden phase transitions
disappear and the calculations predict
a regular increase of the moment of inertia
without any discontinuity (Fig.~\ref{FIG1}).
Still one notes that at \ho=0 the moment of inertia
is smaller than the value suggested by an extrapolation of
the experimental data.
Since we consider this work as a first step towards
a more complete analysis of the effective interaction in
the particle-particle channel, we have
chosen to further renormalize the pairing strength so as to have
a value of $\jde$(0) in agreement with experiment.
This is obtained by the reduction of both pairing intensities
to the values of $\xgt_N=\xgt_Z=12.6$MeV.
In all likelihood, such a modification
lies within the present domain of uncertainty of a pairing force
defined entirely on the basis of static properties such as
quasiparticle energies at ground state or deformation
energy curves. Dynamical properties such as the moment of inertia
$\jde$ will certainly introduce stronger constraints.

With these reduced values of
$\xgt_\tau$, we find that at \ho=0 the neutron and proton
pairing gaps of \Hg{2} are equal to
$\dd_N$=0.557MeV and $\dd_Z$=0.647MeV,
respectively.
The evolution of pairing energies versus {\ho} (Fig.~\ref{FIG2})
shows an interesting
feature: their rate of decrease versus
{\ho} is  reduced together with their values.
This yields
larger pairing correlations at high frequencies and hence
a slower increase of $\jde$ versus {\ho} in better
agreement with data (Fig.~\ref{FIG1}).
At the highest spins reached in the present study,
the pairing correlations are still present
for all the nuclei considered.
For instance, the pairing energies $\CEP{N}$ and $\CEP{Z}$
in \Pb{4} are at $J_0$=52$\hbar$ still of the
order of 0.5MeV (see Fig.~\ref{FIG12}).

Had we been interested in a purely phenomenological analysis
of data, we could probably have improved the agreement by
introducing values of
$\xgt_N$ different from those of $\xgt_Z$ while keeping the
value of $\jde$(0)
consistent with the experimental data.
On the other hand, in our opinion the relative
importance of the neutron and proton pairing should
be ultimately determined by the isospin symmetry.
Unfortunately, this symmetry cannot be
straightforwardly implemented within a
seniority interaction which is by  defined separately for
neutrons and protons. We have therefore chosen to
defer to further studies such an analysis which will
involve a short-range pairing force.
All the results, discussed hereafter have been obtained
with the the pairing strengths $\xgt_N=\xgt_Z=12.6$MeV.

\subsection{HIGH-SPIN PROPERTIES OF GROUND-STATE SUPERDEFORMED BANDS}

In this section we present results for the ground superdeformed
bands of the \Hg{0}, \Hg{2}, \Hg{4} and \Pb{4} nuclei.
Fig.~\ref{FIG4} shows the evolution of their charge quadrupole
moments $\qp$ versus
\ho. The results for the four nuclei are qualitatively different.
The curves for \Hg{2} and \Hg{4}
display an expected behaviour (see also Fig.~3 in
ref.~\cite{SAT91}):
after a small increase
that can be attributed to the diminishing of pairing correlations,
the usual effect of quantum rotations \cite{TRO80}
takes over
and leads to a decrease of $\qp$.
For \Pb{4}
the deformation of the minimum of the SD well
at $J_0$=0 is larger
than in the mercury isotopes,
even when the 4\% effect on $\qp$ associated with the two
additional neutrons is taken into account.
This can be explained by considering
two characteristics of the $Z$=82 subshell
effect (see Fig.~\ref{FIG3}):
first it is shifted to slightly larger deformation
than the $Z$=80 one; second it is weaker, so that the neutron $N$=112
shell effect acts more efficiently and pulls the deformation to larger values.
Similar properties are visible in the particle proton spectrum
obtained with the WSNS method (see Fig.~6 of ref.~\cite{SAT91}).
Another specificity of the evolution of $\qp$ for \Pb{4}
is its constant growth versus {\ho}.

The curve for \Hg{0} displays two distinct patterns.
At low angular velocities
(\ho$\le$325keV) the behaviour is similar to that of the other two
Hg isotopes. Then occurs a neutron two-quasiparticle
alignment (see below) so that the neutron pairing
is reduced (see Fig.~\ref{FIG12}). This
alignment leads to a further reduction
of the quadrupole moment  as
discussed in \cite{TRO80}.

These calculated variations of $\qp$
are smaller than the present level of accuracy of the experiments
which are still consistent with unchanging quadrupole moments
\cite{MOO90,WIL93}.
We hope that coming improvements in the $\gamma$-ray detectors
will allow the observation of such fine differences
in the behaviour
of the quadrupole moment versus {\ho}.
These variations are however sufficiently large to significantly
modify the dependence of single-particle routhians on {\ho}.
This is so because  deformation effects on single-particle
states are in general  stronger than
those induced by
rotation. Fig.~\ref{FIG13} presents the single-particle
routhians in \Hg{0} calculated along the path of self-consistent
quadrupole moments. One can see that after the
crossing the
{\ho}-dependence of the routhians is modified,
 and more pronounced because of faster
changes of the
deformation.

{}Fig.~\ref{FIG5} shows the single-particle
routhians of \Hg{2}.
There is a large proton
shell effect for $Z=80$ and  weaker neutron
ones for $N$=104, 110, 112 and 118. There exists also a
weaker proton shell effect for $Z$=82. This spectrum
looks much like that obtained within the
 WSNS \cite{RIL90,DRI91}
approach. However, we note  some small deviations
which may explain some of the features that we discuss below.

In the following, we label the single-particle
and quasiparticle routhians by their dominant components
in the asymptotic Nilsson basis. Let us recall that
this labelling is mostly useful to attribute names to
the discussed orbitals. Indeed, the dominant components
have sometimes weights as low as 20\% (see ref.~\cite{MEY92} for a detailed
analysis).

In the neutron spectrum, the {\skm} force places
the intruder orbitals \nsu{}, \nsd{} and \nst{}
about 0.8MeV lower  than does the WSNS approach
(with respect to the rest of the spectrum).
On the other hand, the proton
spectrum is very similar. The only noticeable difference is
the existence in our calculation of a bunching
of three levels at the Fermi surface
which leaves the \psd{} intruder level isolated with
small gaps at $Z$=72 and $Z$=74.
In the WSNS spectrum, these four levels
are more evenly spread over 0.8MeV.
The effect of the
change of deformation between \Hg{0} and \Hg{2} can be
seen by comparing the proton single-particle routhians
in Figs.~\ref{FIG13} and \ref{FIG5}.
At $J_0$=0 the three orbitals below
the Fermi level are more spread in \Hg{0} than in \Hg{2}
and the $Z$=74 gap is less visible.
This spectrum is therefore
more similar to that in ref.~\cite{SAT91}.

{}Figure \ref{FIG6} shows the evolution of the
second moment of inertia $\jde$
for \Hg{2}, \Hg{4} and \Pb{4}. The general behaviour of the data
\cite{FAL93a,HAN93,GAL93a,HAN92}
is correctly reproduced,
in particular the apparition of a plateau
at large {\ho} is obtained.
We also reproduce the relative positions of the curves for
the three nuclei:
\Hg{4} lies above  \Hg{2} which itself is above \Pb{4}.
Presently
we have no explanation for the slightly too rapid increase of
the HFBLN moment of inertia.
An analogous deficiency is also present in the recent
WSNS LN calculations \cite{SAT93} for \Hg{4}.
However, since the
results can be
very sensitive
to moderate changes of the pairing force,
we think that it could be
attributed to
a slight imbalance between the neutron and proton strengths.
The fact that we predict a too early occurrence of the plateau for
\Hg{2} and \Hg{4} and the right one for \Pb{4}
supports such a conjecture.

Our calculations show that the values of $\jde$ are
not simply related to the deformation,
contrary to what would occure for
deformed rigid bodies.
For instance, the nucleus \Pb{4}
which, among the studied nuclei, has the largest quadrupole
moment,  has the smallest moment of inertia.
This shows to what extent the magnitude of a dynamic quantity
such as $\jde$ is unrelated with  a static
property such as $\qp$.
Our finding for \Pb{4} is also in agreement with the general
observation that for SD nuclei the rate of
$\jde$ variation versus {\ho} is correlated with the
occupation of the intruder orbital \pst{}.

The calculated and experimental \cite{DRI91}
second moments of inertia of \Hg{0}
(Fig.~\ref{FIG7}) grow also slowly
as a function of the angular momentum until
\ho=325keV where the calculation gives a peak due to
the
two-quasiparticle
alignment associated with the \nsd{} orbital,
see Fig.~\ref{FIG14}. Note that
it is precisely at this frequency that
the single-particle
routhian\footnote{
The superscripts $+$, $-$ or $\pm$ at the Nilsson labels
denote the signatures of orbitals}
\nsd{-} crosses  the Fermi level (Fig.~\ref{FIG13}).
This illustrates the fact that in the HFB theory the occupation
probabilities are not directly determined by the eigenvalues
of the mean-field hamiltonian.
Although the experimental error is still large,
the last point of the
data suggests that a crossing will happen
at a slightly higher angular velocity.

\subsection{QUASIPARTICLE ROUTHIANS IN THE HG REGION}

In the nuclei which differ from \Hg{2} by no more than two nucleon
units, 30 SD bands have already been observed (in
\Au{1}, \Hg{0-4}, \Tl{2-4} and \Pb{4}
\cite{LIA92,LAU92,VOK93,FAL93a,JOY93a,%
HAN93,GAL93a,JOY93c,PIL93,DUP93b,CED93,HAN92}
{}From this information
some general features have already become apparent.\footnote{
Because this is a rapidly evolving domain of physics,
there is unfortunately no consensus on the labeling of the bands
in the available literature.
Sometimes signature partners are denoted with consecutive numbers,
sometimes with the same number and alphabetical (a and b) indices.
We shall therefore avoid such labeling and when necessary,
refer to the bands by what seems to be their most
probable structure according to today's models.
}

\begin{itemize}

\item It appears that pairing correlations are always playing a
significant role.
Except for two bands in \Tl{2}, the $\jde$ of all
considered SD bands is growing.

\item The  nuclei having a
comparable content of intruder states have similar
slopes of $\jde$ as function of \ho.
The fact that this is independent of
whether the nuclei are odd or even is an
additional indication that dynamic
pairing correlations (as described
for instance by particle number projection) are present
and are at least as important as the static ones  contained
in a simple HFB description.

\item As in the rare earth region, there are
also specific trends associated
with the content of the proton or neutron intruder-state
configuration. For instance,
one can group several SD bands
into a ``family'' associated
with the ground SD band in \Hg{2}.
These bands have the fastest growing
second moment of inertia $\jde$. Among them there are
several examples of pairs of identical bands.

\item Recently another family associated
with two SD bands in \Tl{3} has emerged \cite{DUP93b}.
It includes mostly bands
assumed to have the \pst{} orbital blocked.
One can also add
to this family the ground SD band in \Hg{1}
which is interpreted as having
the \nsd{} orbital blocked.
If this orbital assignments are
correct, this would indicate
that the
effects of proton and neutron intruder states
are of a similar magnitude.

\item A few bands, one recently discovered in
\Hg{3} \cite{JOY93c} and
two in \Tl{2} \cite{LIA92,DUP93b}, display a markedly different
behaviour. The latter two can be explained by
a blocking of two (neutron and
proton) intruder states at the same time.
Similar bands have also been observed in \Tl{5}
\cite{DUP93b}.
\end{itemize}

We first consider
the HFBLN quasiparticle routhians for \Hg{2}
(Figs.~\ref{FIG8} and \ref{FIG9}). Because we are
going to discuss results for neighbouring nuclei, we have labeled
the routhians (at \ho=0) not only with their dominant Nilsson
configurations
but also with a letter p or h to indicate whether they
are built on a particle or a hole configuration. If a nucleus is
related to \Hg{2} by an addition of
particles, the Fermi level will raise. One expects that
the qpr's
built on hole states will be pushed up while those built on
particle states
will go down. The opposite effect will occur
when the nucleus is obtained
by a subtraction of nucleons.

In the proton diagram (Fig.~\ref{FIG8}),
the qpr associated with the \pst{} orbital
is well separated from the other ones.
It displays a rather gentle
alignment pattern and
a very small signature splitting. This is in agreement
with the observation that in all Tl isotopes, the bands can be sorted
in pairs of signature partners. Moreover, the fact that
their moment of inertia $\jde$
is always growing less rapidly than that of \Hg{2} is
consistent with having one intruder orbital blocked.

Experimental routhians can be extracted from
the information for the two bands of \Tl{3}
by taking \Hg{2}
as a reference and making reasonable assumptions for the spins.
They show a decrease
of this particular qpr of about 0.4MeV in the range
of angular velocities of 110$\le$\ho$\le$360keV
and a maximal signature splitting of 30keV. In our calculations
we find
0.6MeV  and 80keV, respectively,
for these two quantities.
A similar analysis of the
\pst{} qpr in \Pb{4} leads to smaller numbers: 0.52MeV and 50keV.
It therefore seems that
our calculations predict a slightly too large
proton alignment. This
could explain why the HFBLN moments of inertia (Fig.~\ref{FIG6})
increase somewhat faster than
the experimental data.

The qpr's associated with
the three hole levels \pctdt{}, \pctzu{} and \pquuu{}
are located around 1MeV.
{}For lighter neighbouring elements,
these qpr's  become the lowest
ones and replace in the qpr spectrum
those built on the particle orbital \pst{}.
In \Au{1} our calculation would therefore predict three bands
close in energy. The one built on \pquuu{} would not have an easily
detectable partner because of the strong signature splitting.
It is tempting to say that this is the band which has
already been observed.
The other two result from a
strong interaction between the levels
\pctdt{} and \pctzu{}, which pushes down the
qpr's \pctdt{+} and \pctzu{-} of opposite signatures.

In this respect, our calculation differs slightly from that using
the WSNS method. Indeed, because the particle routhian
built on the
orbital \pctdt{} is lower in the spectrum of ref.~\cite{VOK93},
there is less interaction with the
one built on \pctzu{}. Since this orbital
has a large signature splitting, one should then in principle
expect to find only one other band without a partner.
However, we note that the signature splitting of levels
(see for instance \pquuu{} in Fig.~3 of ref.~\cite{VOK93})
in the WSNS approach is smaller than the one found in the present work.
Therefore their predicted
signature partners are never very far in energy.
In addition, the negative-signature intruder
orbital \psd{-} obtained in ref.~\cite{VOK93}
seems to interact earlier (\ho$\approx$250keV) with
\pquuu{-}. It crosses it and appears at the Fermi surface
at \ho=0.300keV.
In our calculation, until \ho=410keV, the \psd{-} state is still not favoured
as compared to the orbitals \pquuu{-} and \pctdt{\pm}.
Further experiments on gold isotopes may certainly help
clarify this situation.

Let us now consider the  qpr spectrum for neutrons (Fig.~\ref{FIG9}).
At $J_0$=0 one first finds a group of three levels.
Two are of the  particle type, \ncudc{} and \nsdqn{}, and one of the hole
type \nsqdt{}. Further up there is a group of four levels. Among them
there are two routhians built on intruder orbitals: the
particle-type
\nst{} and the hole-type \nsd{}. Because of a strong interaction
between these two intruder routhians,
they almost immediately exchange their character
when {\ho} increases. In particular, it seems more
appropriate to assign the steep downgoing branch which comes out of
the qpr labeled \nst{} to the orbital \nsd{}. At this point,
we will only justify this affirmation
by noting that the signature splitting of the latter
routhian is larger
than that of the former one (see Fig.~\ref{FIG5})
because it has a smaller value of $K$.
Later, we will
analyze this point
in more details
when comparing the neutron qpr spectra of the four mercury
isotopes studied in this work.

If we try to extrapolate from this  spectrum
that of \Hg{1}, we expect
that the most interesting configurations will be those associated
with the
hole-type qpr's: those of the
two signature-partner
states of \nsqdt{} and the negative-signature
down-going state \nsd{-}.
We note that for \ho$>$300keV the latter qpr is the least
excited one.
Moreover, in \Hg{0} the same state is the least excited
already at \ho=120keV (Fig.~\ref{FIG14}).
One can therefore
infer that the
corresponding crossing will happen in \Hg{1}
at an intermediate angular velocity.
Since the SD bands are essentially fed at high spins,
it is natural to
expect the \nsd{-} band to be the most populated.
Then should come two bands with moderate signature splitting
associated with \nsqdt{\pm}. This corresponds well to the experimental
situation. The extraction of the experimental qpr's for the
two signature-partner bands shows an
approximate overall decrease of 150keV with a
maximal splitting of about 40keV over the range
of 150$\le$\ho$\le$380keV.
For these two numbers we find 330keV and 100keV, respectively.
Therefore, as was the case
for protons
in \Tl{3},  we find
also a neutron alignment which is
 slightly too large.

We consider now the heavier isotope \Hg{3}.
A hole state such as \nsqdt{} will now become higher than the
two particle states \ncudc{} and \nsdqn{}.
On the latter orbital two
SD bands with a negligible signature splitting
can be built. Candidates for these bands
exist in the experimental data.
Similarly, there should also be a band built
on the \ncudc{+} qpr. This band should be very similar to the
two previous ones because of the strong similarity of
their qpr's
which remain almost parallel to each other.
This has also been observed.
In addition our diagram shows that the negative-signature
orbital \ncudc{-} interacts strongly with \nsd{-}.
Such a phenomenon has precisely
been seen in the experiment \cite{CUL90,FAL93a,JOY93a}.
Moreover, recently a candidate for the \nsd{+}
band has been observed \cite{JOY93c}
with a reduced $\jde$ which is compatible with an
upbending behaviour of the associated qpr, as is visible
in our qpr spectrum. The {\ho} value for the crossing is
measured to be equal to 250keV and compares favourably with
the value of 200keV obtained in our calculations.

Let us now pursue this discussion by
considering in more details the evolution of the
neutron qpr spectrum as function of the neutron number
for the mercury isotopes.
{}Fig.~\ref{FIG10} includes selections of
levels from the qpr spectra obtained for the
ground SD bands of \Hg{0}, \Hg{2} and
\Hg{4}. We have also included the spectrum obtained by a
calculation of \Hg{3} treated as an {\em even} nucleus,
using the fact that the number of particle of a BCS wave-function
is only defined on the average. The qpr's shown in Fig.~\ref{FIG10} for
\Hg{3} result from a calculation
in which we have arbitrarily imposed that the average neutron number
is equal to 113.
This allows us to study the influence of
the position of the Fermi level on the crossing of the
negative-signature
\ncudc{} states with the $N$=7 intruder states.

The qpr's shown in Fig.~\ref{FIG10}
are \nsqdt{} and \nsd{} for the holes states, and
\ncudc{} and
\nst{} for the particle states. To make the figure more readable,
we have eliminated the qpr's associated with
the hole states \nczco{} and \nsqzu{} which play no role in our discussion and
the particle state \nsdqn{}. The latter qpr shows
no sign of  signature splitting. In all
Hg isotopes, it remains parallel to
the \ncudc{+} orbital with an energy shift which
decreases gradually from 150keV in \Hg{0} to almost zero
in \Hg{4}.
As mentioned above, two signature-partner bands in \Hg{3}
have been assigned to this particular orbital \nsdqn{}.

The \Hg{0} qpr diagram shows that the single-particle
dynamics is
determined mostly by the alignment of the strongly
signature-split orbital \nsd{} whose
negative-signature partner crosses the \nsqdt{}
at \ho=120keV, then becomes negative at \ho=300keV and
generates an alignment at \ho=325keV.
At this frequency, the experimental $\jde$ indicates no
alignment
but the last available data point \cite{DRI91} suggests
a possible crossing at a higher angular frequency.
Moreover, we obtain a
slightly too large slope of the second moment of inertia
prior to the peak. Both facts suggest again that
our calculations predicts an alignment faster than seen in the
experiment.

{}From the separation of the qpr's, we estimate that the interaction
between the ground SD band and the
two-quasiparticle band is approximately
300keV.
According to our diagram it seems also that
a negative-parity band built on the \nsqdt{-} will become yrast
beyond \ho=400keV.

In the qpr diagrams of \Hg{2}, \Hg{3} and \Hg{4}
(Fig.~\ref{FIG10}), there is a modulation of the interaction
between the intruders \nsd{-} and \nst{-},
and the \ncudc{-} orbital
as the hole-type qpr's
move up in the spectrum. In \Hg{4}, the \nsd{} is pushed so high
by the displacement of the neutron Fermi level that despite its
rapid descent, its negative-signature branch
\nsd{-} does not interact
with \nst{-} before \ho=360keV.
Had we pursued the calculation to higher
spins, an extrapolation suggests that an interaction
with the \ncudc{-} would have occurred near \ho=440keV. It is interesting
to see that
in these four isotopes of mercury,
the \nst{+} curve displays precisely
the upbending behaviour of the experimental routhian
of the band recently discovered in \Hg{3} \cite{JOY93c}.
Qualitatively speaking, one could then say that the SD band interaction
observed in \Hg{3} concerns the \nsd{-} and \ncudc{-} states
while the last SD band in the same nucleus is built on the
\nst{+} state. Of course, we must keep in mind that
within the energy range of the order of the
pairing gap
around the Fermi level,
one cannot distinguish between  holes and particles. As shown
in the qpr spectrum of \Hg{0} (Fig.~\ref{FIG14})
the \ho=0 identity of the qpr's is rapidly lost if they
happen to be close in energy and have large Coriolis coupling
matrix elements, as is the case for \nst{} and \nsd{}.

The qpr spectrum of \Hg{3} is intermediate between that of
\Hg{2} and \Hg{4}. One sees that the lowering of the \nsd{}
is not large enough to induce an interaction at the values of
{\ho} where it has been observed. In our calculation
the interaction only occurs at \ho=380keV.
Nevertheless, there does not seem to be any need
to invoke
dynamical octupole correlations to explain the value of {\ho}
at which the band
crossing occurs in \Hg{3} \cite{RIL90}.
This is also in line with the results of the
Generator Coordinate Calculations \cite{BON91b,BON93a}
which find that a collective one-octupole-phonon SD band
would be excited by approximately 1.7MeV with respect to the
ground SD band, and therefore difficult to detect with presently
available techniques.
In fact, a mere 200keV up shift of
the $J_0$=0 position of the \nsd{} intruder state would certainly
eliminate most of the discrepancies between our calculations
and experiment.
We will come back to this question in the concluding section.

The neutron qpr spectrum of \Pb{4} is presented
in Fig.~\ref{FIG11}.
A comparison with  the
isotone nucleus \Hg{2} (Fig.~\ref{FIG9}) shows the influence
of the proton-induced polarization.
According to the above discussion,
because the $Z$=82 SD shell effect is weaker than the $Z$=80 one,
the deformation of \Pb{4} is larger.
As a result (see for instance Fig.~\ref{FIG3}),
the \nst{} orbital is approximately equidistant to
\nsd{} and \ncudc{} at \ho=0 (in \Hg{2}, it
is closer to \nsd{}). Now instead of two successive interactions,
one observes a global interaction involving the three
negative-signature
levels for values of {\ho} ranging from 150keV to 450keV.
This behaviour so different from that observed in \Hg{2}
results from a mere 50keV shift of the \nst{} qpr away from the
\nsd{} orbital.

\section{Summary and conclusions}

In order to analyze nuclear structure at high angular momentum, we have
constructed a cranked Hartree-Fock-Bogoliubov code. We solve the
equations iteratively by performing in sequence the following
operations.
In the first one we evolve the single-particle wave
functions by one imaginary-time step using the
cranked Hartree-Fock mean field. In the second one we
build the HFB matrix and solve the
associated eigenvalue problem. After selecting the
occupied quasiparticle states
we compute the one-body density matrix and the pairing tensor. Then the
eigenfunctions of the density matrix and the associated
occupation probabilities are used to define the
potential and the effective mass which enter the
cranked Hartree-Fock
mean field. This completes one iteration.

This method of solution which requires the simultaneous handling of two
single-particle bases, allows an easier separation of the particle-hole
and particle-particle (pairing) channels.
It also permits a significant reduction of the size of the HFB problem.
Since it is our feeling that the mean-field part of
the effective force is reasonably well known (at least in the vicinity
of the stability line), we plan to use this method
to explore in more detail the pairing channel whose influence
on the rotational properties of nuclei is expected to be
important.

A HFB test calculation for the nucleus \Hg{2} has shown the limitations
of the mean-field method. At the angular velocities for which either
proton or neutron pairing  disappears , the dynamical
moment of inertia displays abrupt decreases which are not compatible
with the experiment. It is known that this deficiency results from
a poor treatment of pairing correlations by the mean-field approach in the
critical and subcritical regimes. It is also known that this can be
improved by a variation after particle number projection (VAP). We have used
the Lipkin-Nogami prescription (HFBLN) which provides a good approximation
of the exact VAP. As expected, the HFBLN results for $\jde$ do not display
spurious phase transitions and are in good qualitative agreement with data.

We have studied the ground superdeformed bands of the four nuclei
\Hg{0},
\Hg{2}, \Hg{4} and \Pb{4}. The agreement is generally good. Using
a single parameter for the strength of the pairing force,
and adjusting it to reproduce the
experimental dynamical moment of
inertia at low frequency,
we can obtain the slow upgoing
behaviour of data including the appearance
of a plateau around \ho=300keV. We also find that the $\jde$ of \Pb{4}
is smaller than that of \Hg{2} which itself is smaller than that of \Hg{4}.
An analysis of the quasiparticle routhians
has shown that our spectrum
is compatible with the single-particle structure of the SD well as it
can be deduced from the present experimental
knowledge based on 30 bands in nuclei which are close
neighbours of \Hg{2}. This includes the interesting band-interaction effect
observed in \Hg{3}.

Despite this good agreement, we have detected some weaknesses in our results.
The most evident one is a slightly too fast alignment.
It seems reasonable to assign this property to a
too simple form of the pairing interaction.
Several lines of improvements appear
possible. First one could consider modifying the relative balance of
the neutron and proton strengths. One can also study the effects coming
from a more refined interaction such as for instance a density-dependent
zero-range force.

It has also  become apparent
that a good description of data requires a very precise
determination of the particle spectrum at \ho=0. We have seen that our
spectrum differs from that obtained within the WSNS approach \cite{SAT91}
as far as the positions of some neutron intruder states
are concerned.
The differences between the models may
originate from small
variations on the position of the intruder states at
sphericity. Indeed
any such difference will be emphasized by the large deformation
associated with superdeformation. Another cause might be the
value of the equilibrium deformation at the minimum
of the SD well. Since we are often
dealing with states which
in the Nilsson diagram are rapidly up- or downgoing, a modification
of the equilibrium deformation (of the order of 1e.b for $\qp$) could explain
the differences between our spectrum and that of ref.~\cite{SAT91}.
Since we have seen that the deformation of proton and neutron shell
effects are slightly offset,
a small modification of proton versus neutron pairing
strengths could probably realize this effect.
In fact, the uncertainty as to the exact magnitude of the
deformation is only one more example stressing
the importance of polarization effects on the high spin
behaviour of nuclei: in this work, we have also
discussed a case where a change in proton occupation modifies the behaviour of
the neutron quasiparticle routhians.

Altogether the differences between various approaches
in predicting the high-spin structure remain rather small,
at least for those which also successfully
describe collective properties at zero spin.
Such differences could even be considered as being within the
predictability range of the
methods themselves.
Nevertheless the challenge is there: the quality of present
experimental information has introduced us into an era where much stricter
requirements will be imposed on nuclear structure models.

We gratefully acknowledge interesting discussions
with W.~Satu{\l}a and R.~Wyss.
One of us (H.F.) would like to acknowledge the hospitality
of the National Institute for Nuclear Theory (Seattle)
where this work was completed.
We thank the European Centre
for Theoretical Studies in Nuclear Physics and Related Areas (ECT*)
in Trento for its hospitality and for partial support for this
project.
This research was supported in part by the Polish Committee
for Scientific Research under Contract No. 20450~91~01.

\renewcommand{\topfraction}{0.0}
\renewcommand{\bottomfraction}{0.0}
\renewcommand{\floatpagefraction}{0.0}
\setcounter{topnumber}{20}
\setcounter{bottomnumber}{20}
\setcounter{totalnumber}{20}

\clearpage

\bc
{\large Figure Captions}
\ec

\begin{figure}[ht]
\caption[FIG1]{%
The second moment of inertia $\jde$
of the ground SD band of  \Hg{2}.
The data (open circles) are
taken from ref.~\cite{GAL93a}.
The other curves correspond to the HF (solid line),
HFB with $\xgt_\tau$=15.5MeV
(triangles), HFBLN with $\xgt_\tau$=12.6MeV (solid circles)
and HFBLN with $\xgt_\tau$=14MeV (solid squares) calculations.}
\label{FIG1}
\end{figure}

\begin{figure}[ht]
\caption[FIG2]{%
Pairing contribution to the total energy
of the ground SD band of  \Hg{2}.
The curves correspond to the HFB with $\xgt_\tau$=15.5MeV
(dashes),
HFBLN with $\xgt_\tau$=12.6MeV (solid line)
and HFBLN with $\xgt_\tau$=14.MeV (dots) calculations.}
\label{FIG2}
\end{figure}

\begin{figure}[ht]
\caption[FIG3]{%
Nilsson diagram obtained by a quadrupole
constrained HFBLN calculation for the nucleus \Hg{2}
at \ho=0. The deformation
is measured by the charge quadrupole moment in e$\cdot$b. The levels are
indexed by their dominant asymptotic configuration. Positive- and
negative-parity levels are drawn as solid and dot-dashed lines, respectively.
The dotted line indicates
the Fermi level. The arrows indicate the charge quadrupole moment of the
SD minimum.
}
\label{FIG3}
\end{figure}

\begin{figure}[ht]
\caption[FIG12]{%
Proton (dashes) and neutron (solid line)
pairing energies in MeV
versus angular velocity for the ground SD bands.
}
\label{FIG12}
\end{figure}

\begin{figure}[ht]
\caption[FIG4]{%
Charge quadrupole moments in e$\cdot$b
versus angular velocity for the ground SD bands.
}
\label{FIG4}
\end{figure}

\begin{figure}[ht]
\caption[FIG13]{%
Single-particle routhians as function
of the angular velocity for the ground SD band of the
nucleus \Hg{0}.
The convention for the different (parity, signature)
combinations is: $(+,+)$ solid line, $(+,-)$ dashed line,
$(-,+)$ dot-dashed line, $(-,-)$ dotted line.
}
\label{FIG13}
\end{figure}

\begin{figure}[ht]
\caption[FIG5]{%
Same as in Fig.~\ref{FIG13} for \Hg{2}.
}
\label{FIG5}
\end{figure}

\begin{figure}[ht]
\caption[FIG6]{%
The second moments of inertia $\jde$ of the
ground SD bands of
\Hg{4}, \Hg{2} and \Pb{4}.
Data for \Hg{4} \cite{RIL90}, \Hg{2} \cite{GAL93a} and
\Pb{4} \cite{HAN93}
are indicated respectively by empty squares, circles and triangles. HFBLN
results are drawn as dashed (\Hg{4}), solid (\Hg{2}) and
dot-dashed (\Pb{4}) curves and solid symbols.
}
\label{FIG6}
\end{figure}

\begin{figure}[ht]
\caption[FIG7]{%
The second moment of inertia $\jde$
of the ground SD band of the nucleus
\Hg{0}.
Data \cite{DRI91} are indicated by open squares.  HFBLN
results are drawn as a solid curve.
}\label{FIG7}
\end{figure}

\begin{figure}[ht]
\caption[FIG14]{%
Neutron quasiparticle routhians as function
of the angular velocity for the ground SD band of the
nucleus \Hg{0}.
The convention for the different (parity, signature)
combinations is: $(+,+)$ solid line, $(+,-)$ dashed line,
$(-,+)$ dot-dashed line, $(-,-)$ dotted line.
}
\label{FIG14}
\end{figure}

\begin{figure}[ht]
\caption[FIG8]{%
Same as in Fig.~\ref{FIG14} for protons in \Hg{2}
}
\label{FIG8}
\end{figure}

\begin{figure}[ht]
\caption[FIG9]{%
Same as in Fig.~\ref{FIG14} for \Hg{2}.
}
\label{FIG9}
\end{figure}

\begin{figure}[ht]
\caption[FIG10]{%
Comparison of the evolution
of selected neutron quasiparticle routhians as function
of the angular velocity for the ground
SD bands of the nuclei \Hg{0},
\Hg{2}, \Hg{3} (see text) and \Hg{4}.
The convention for the different (parity, signature)
combinations is the same as in Fig.~\ref{FIG14}.
}
\label{FIG10}
\end{figure}

\begin{figure}[ht]
\caption[FIG11]{%
Same as in Fig.~\ref{FIG14} for \Pb{4}.
}
\label{FIG11}
\end{figure}

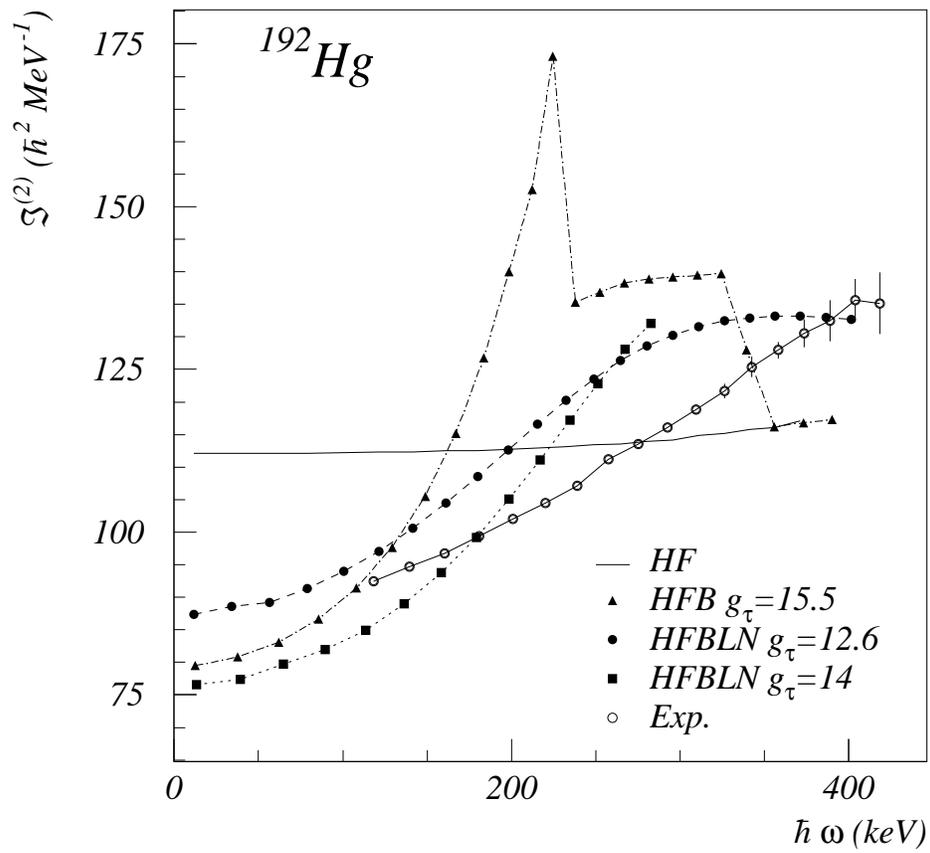
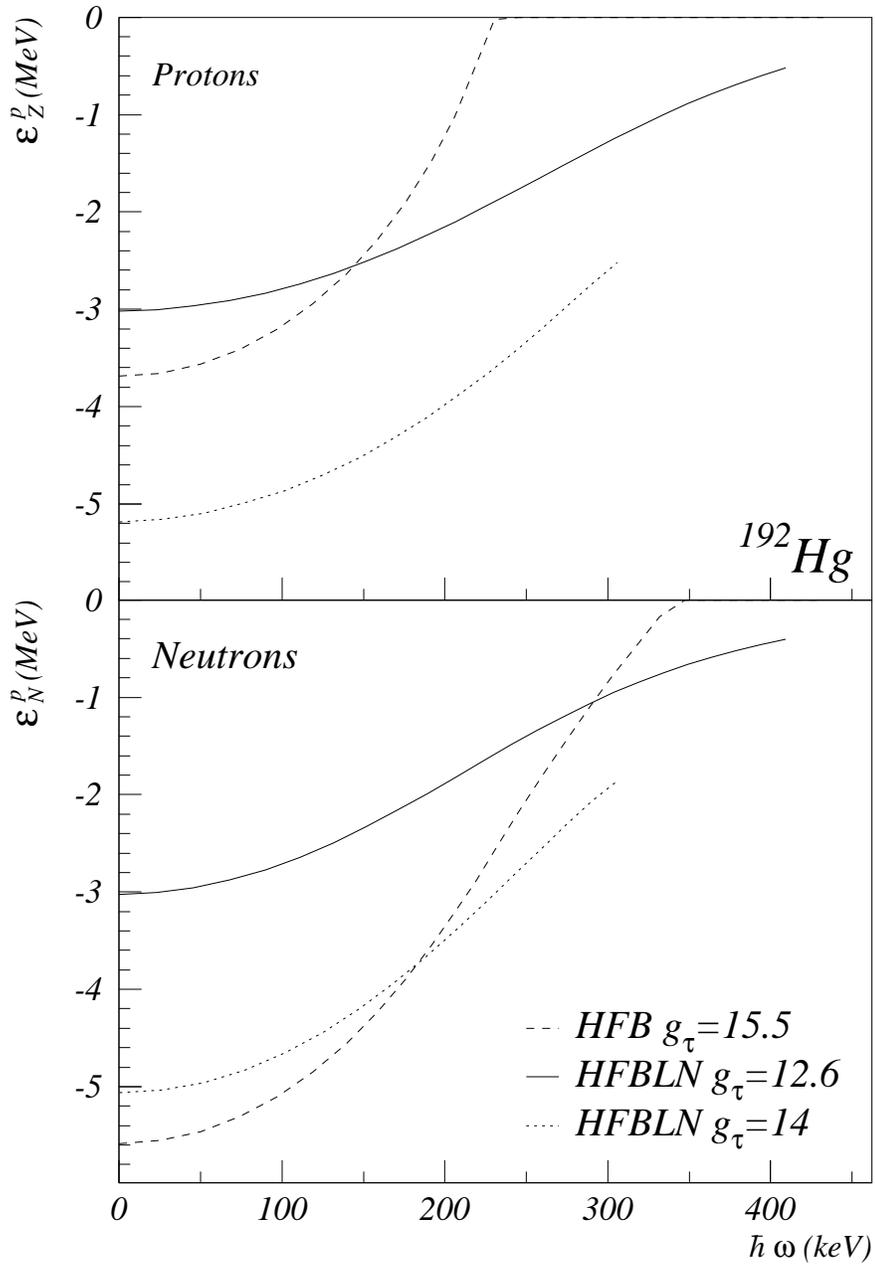
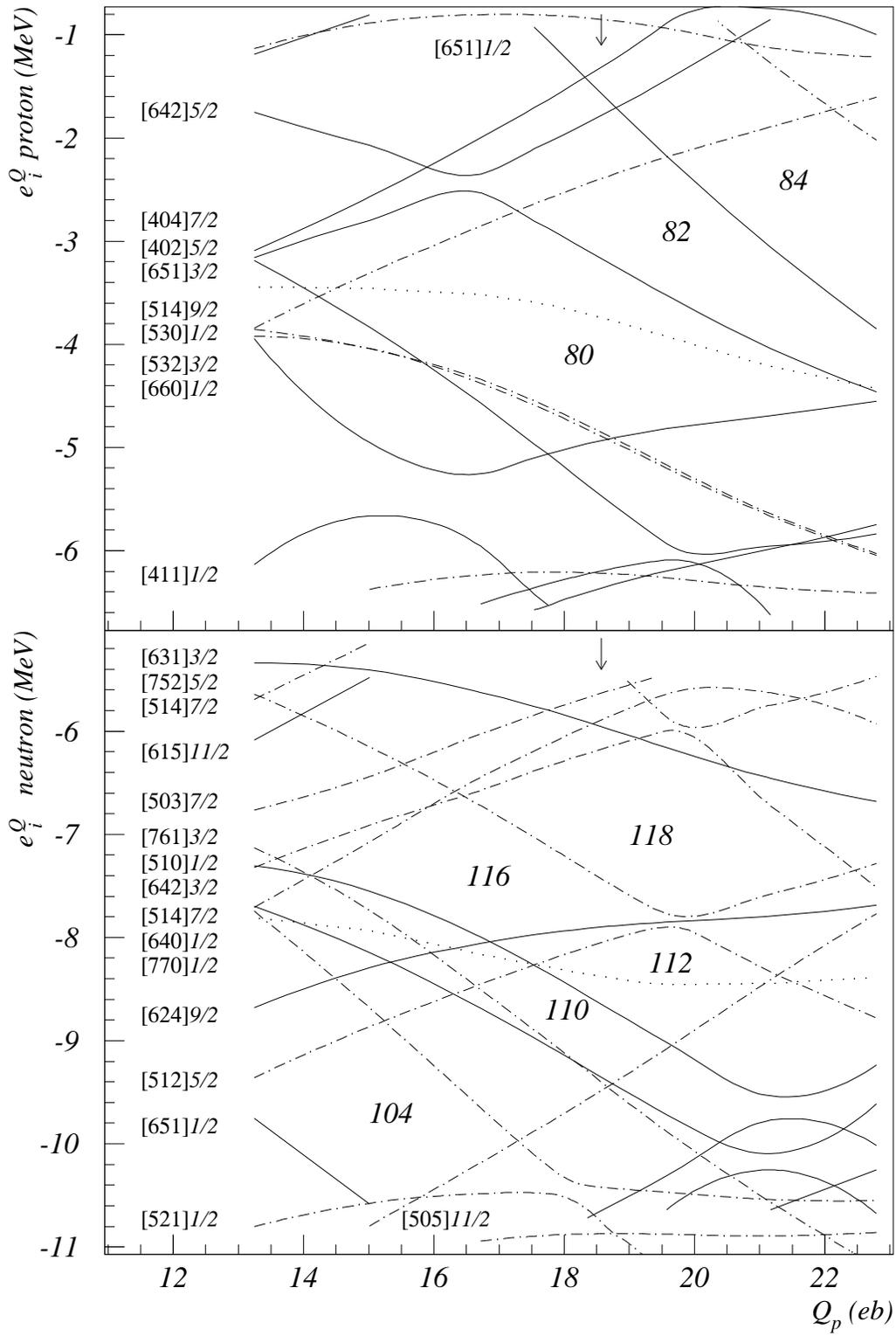
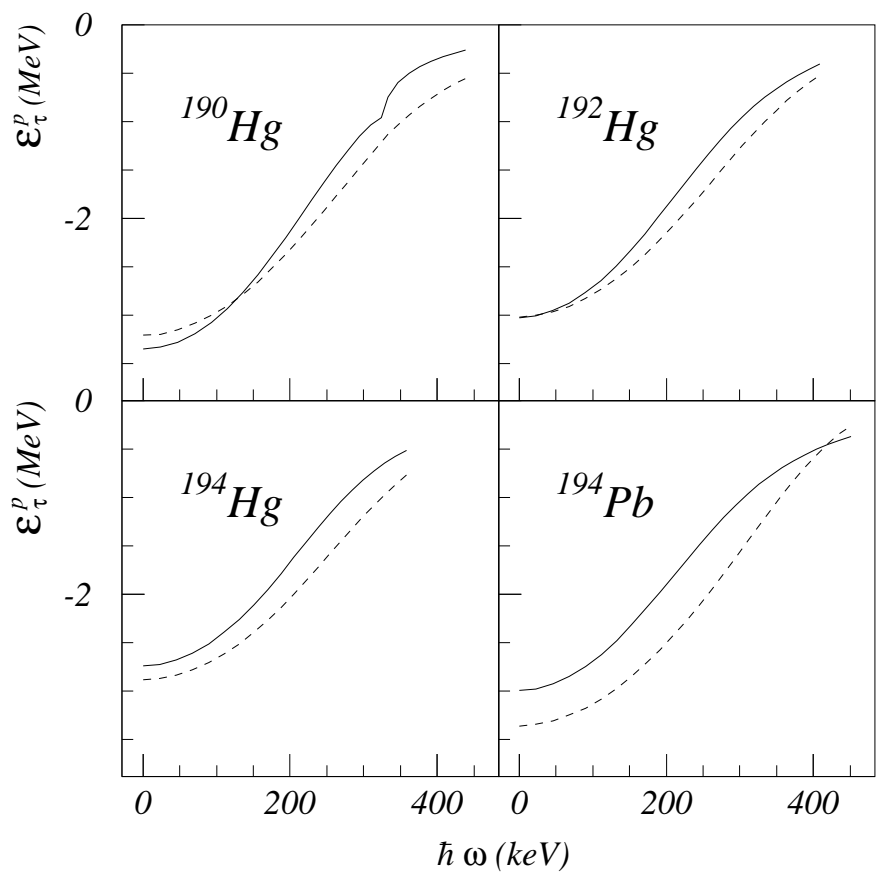
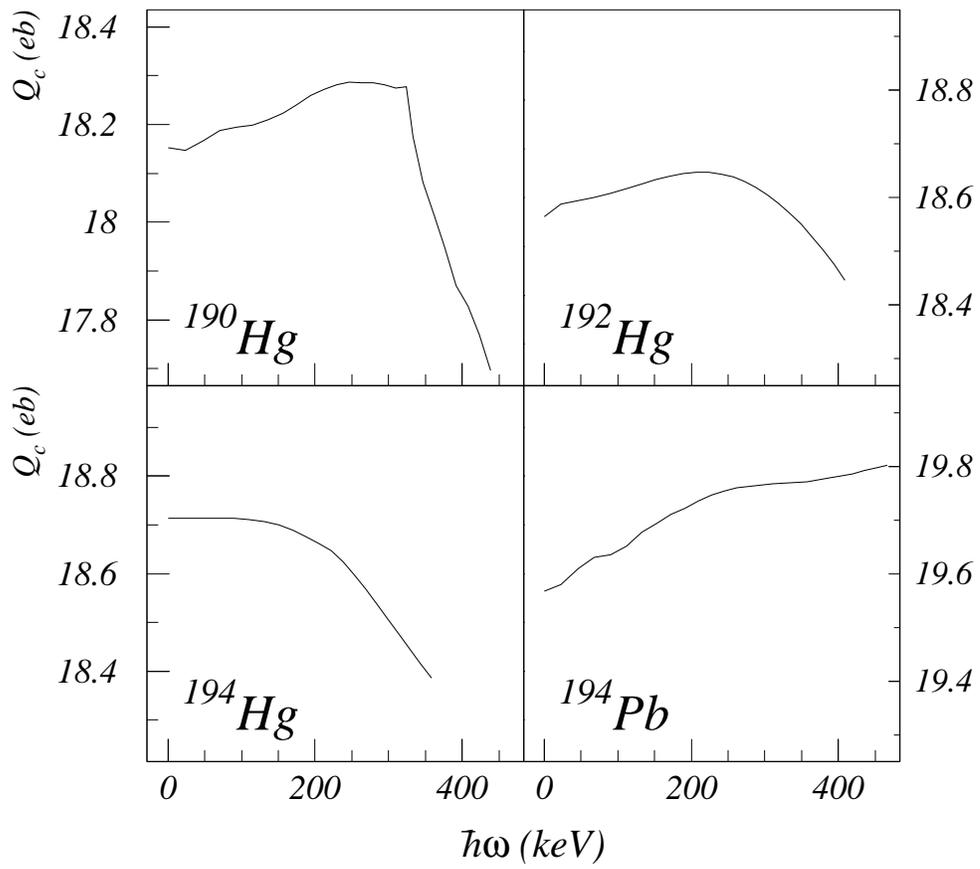
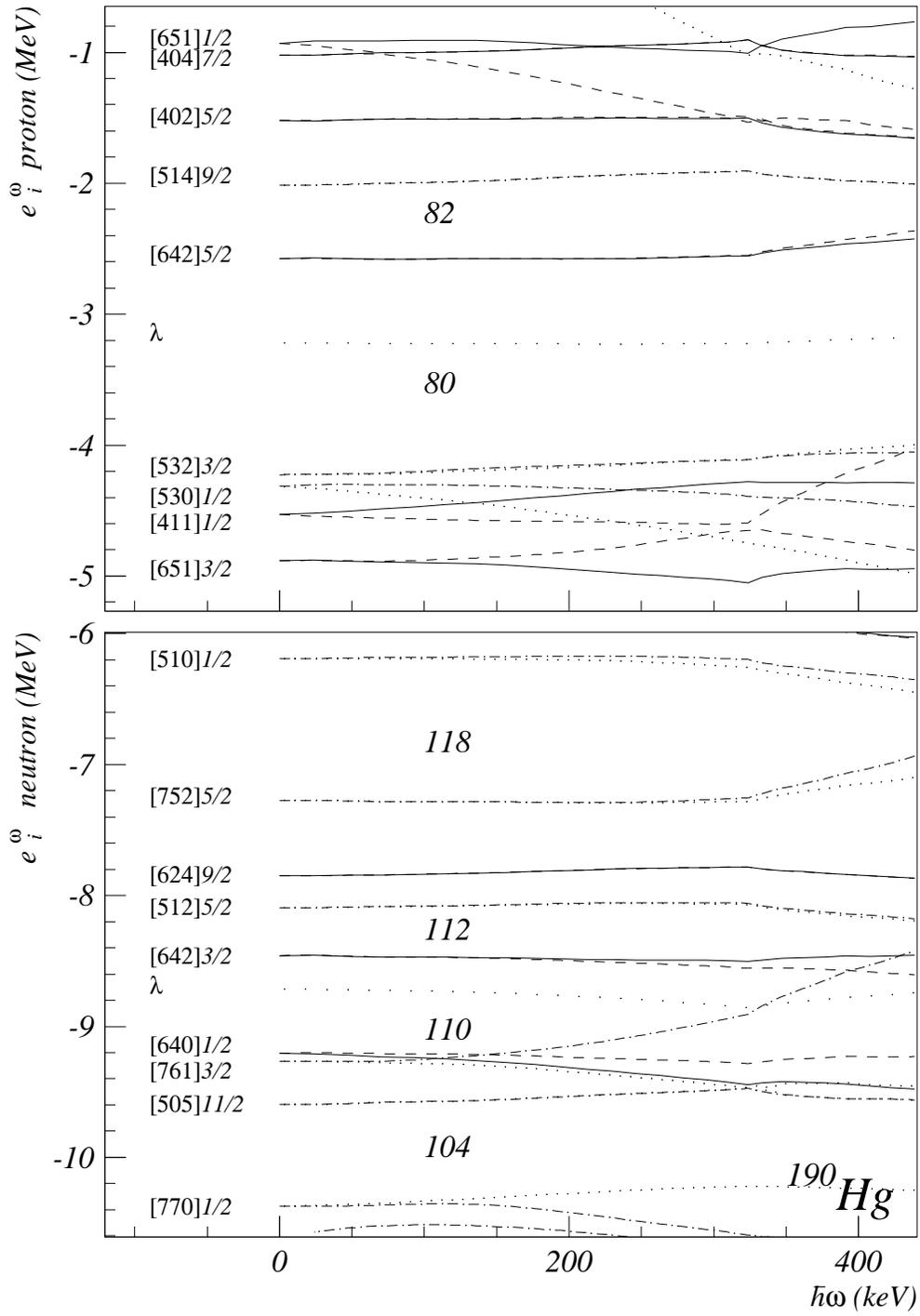
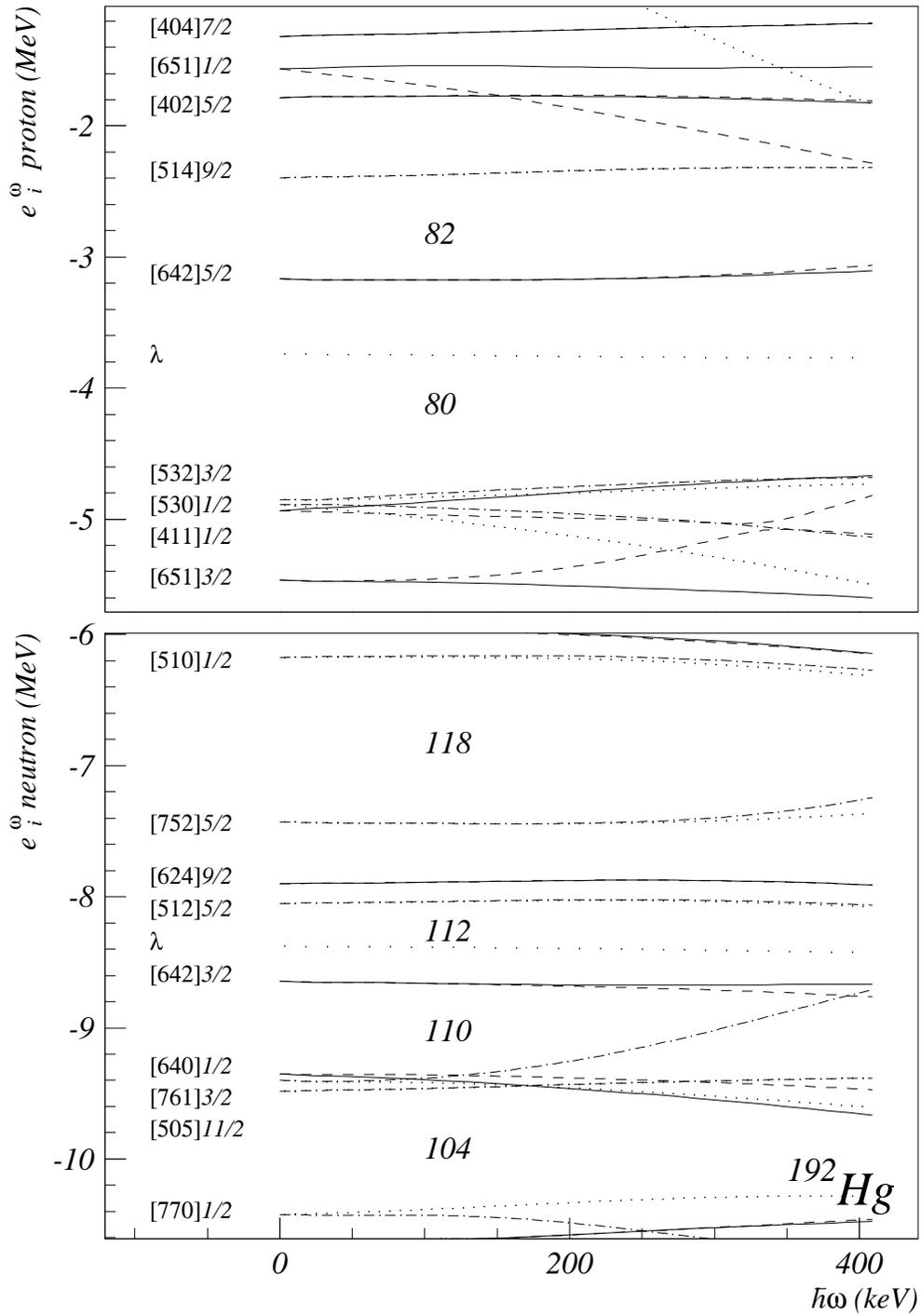
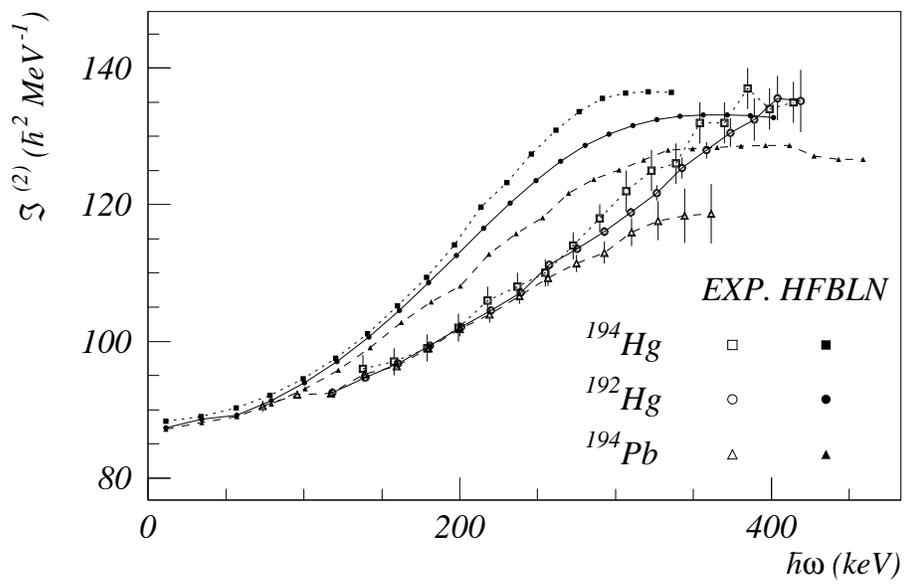
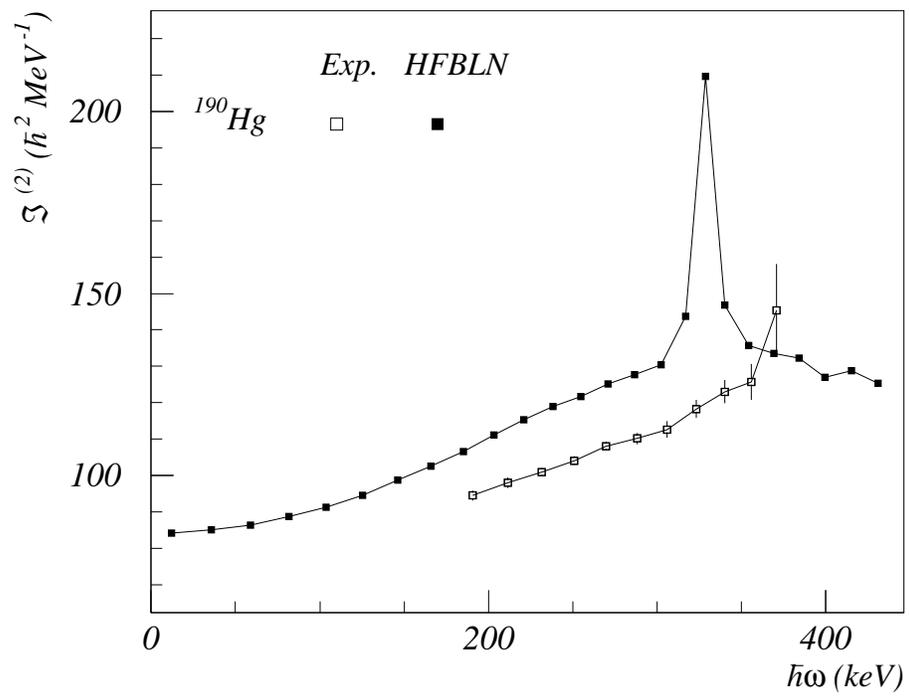
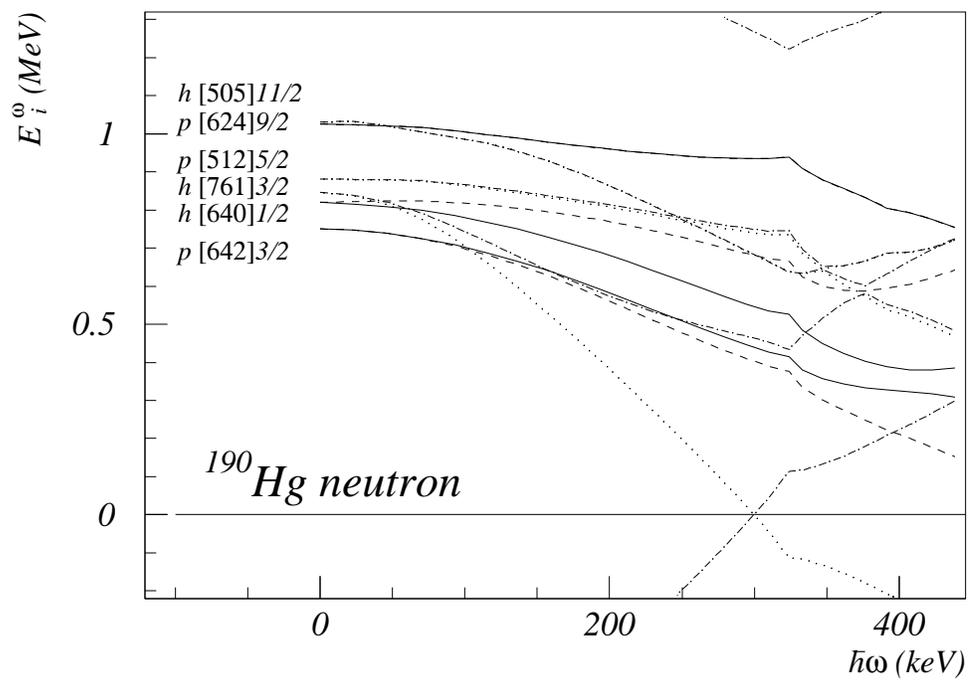
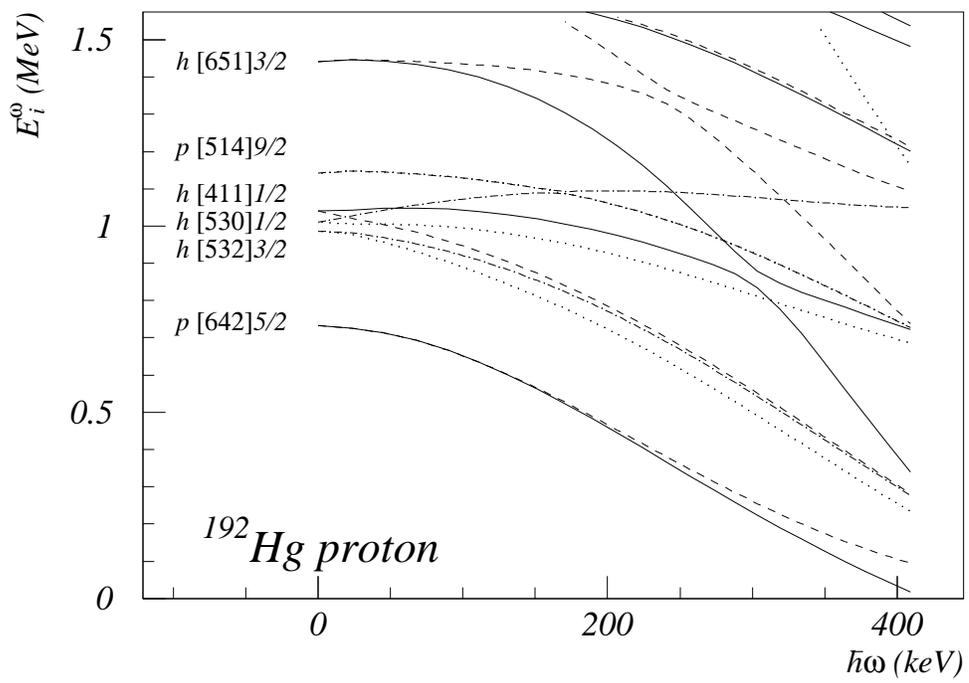
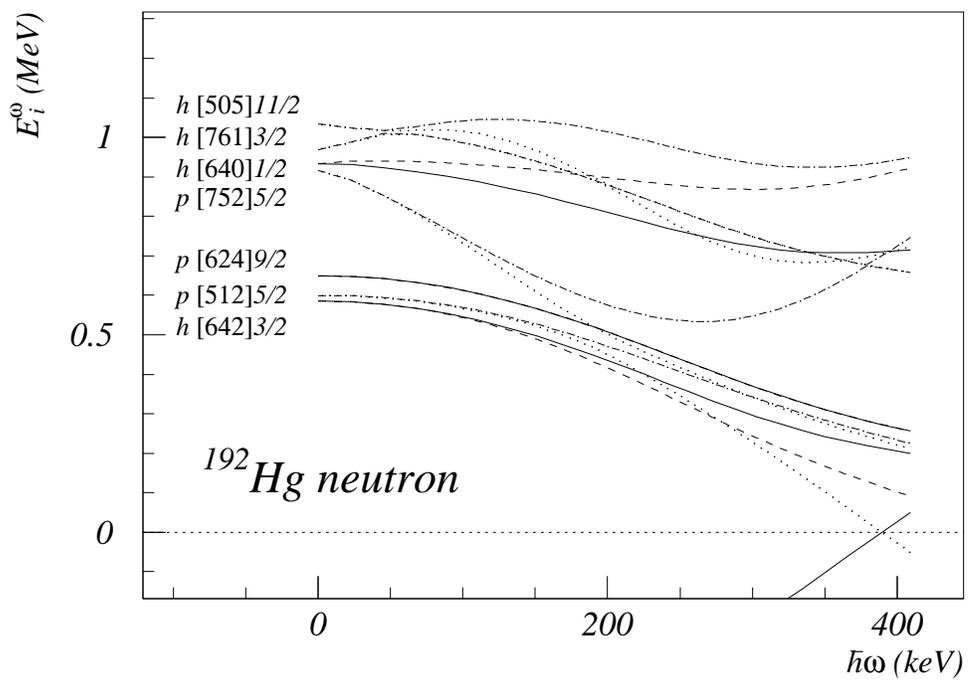
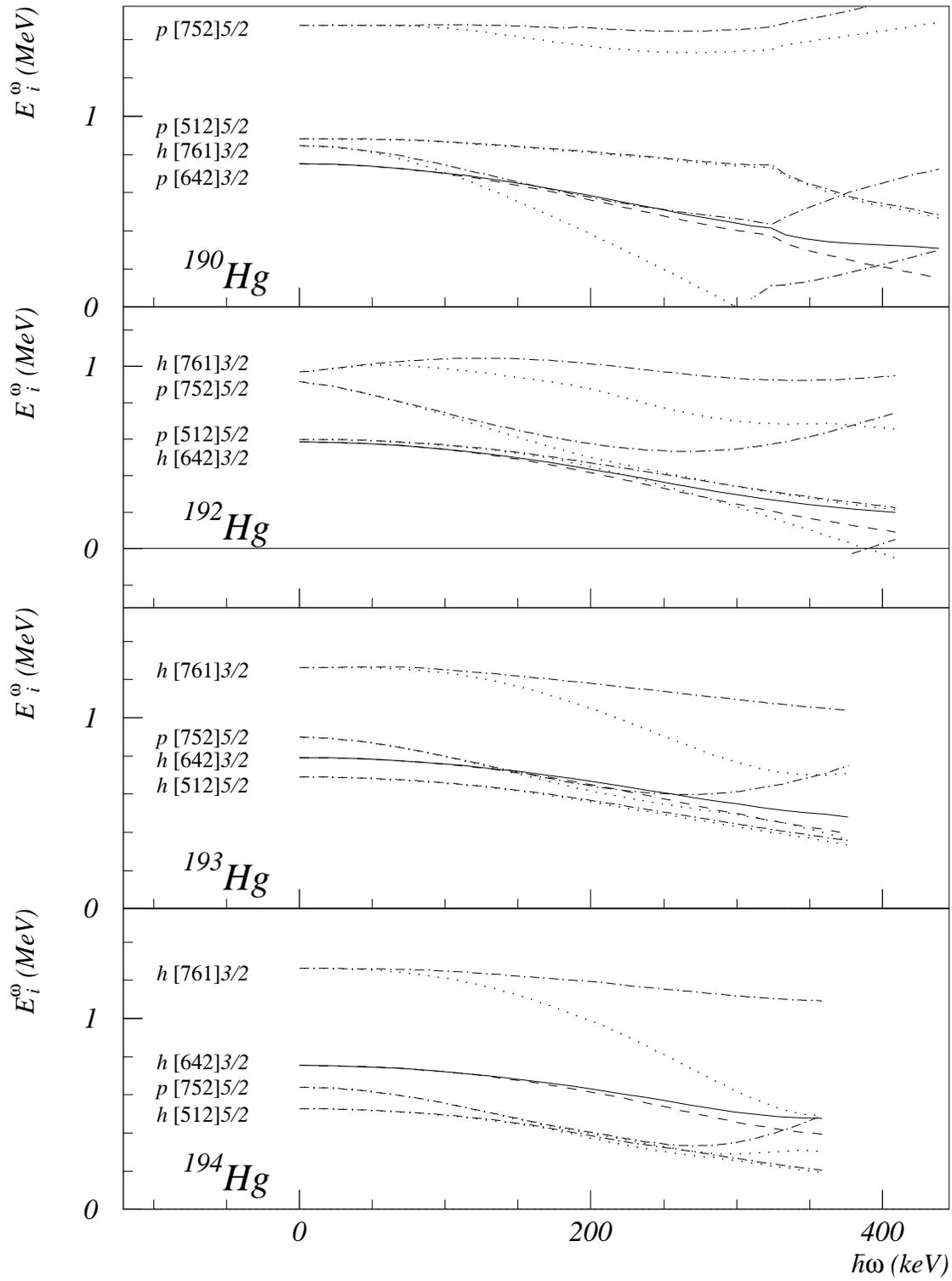
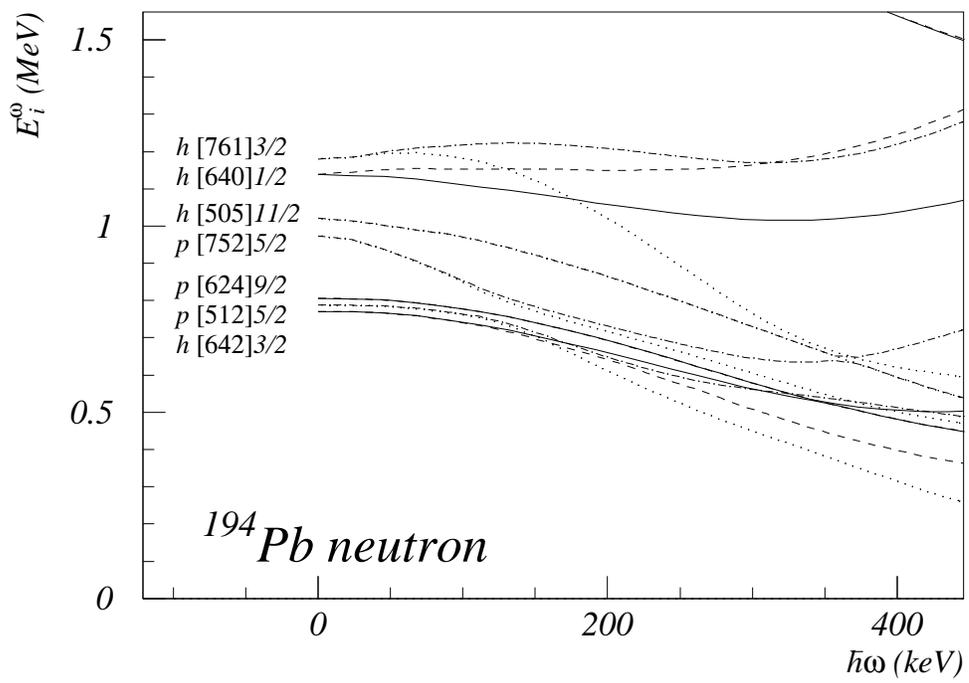
\end{document}